# DOUBLE-HELICAL TILED CHAIN STRUCTURE OF THE TWIST-BEND LIQUID CRYSTAL PHASE IN CB7CB

Michael R. Tuchband,[1,*] Min Shuai,[1,*] Keri A. Graber,[1] Dong Chen,[1] Chenhui Zhu,[2] Leo Radzihovsky,[1] Arthur Klittnick,[1] Lee Foley,[3] Alyssa Scarbrough,[3] Jan H. Porada,[3] Mark Moran,[3] Joseph Yelk,[1] Dmitry Bedrov,[4] Eva Korblova,[3] David M. Walba,[3] Alexander Hexemer,[2] Joseph E. Maclennan,[1] Matthew A. Glaser,[1] and Noel A. Clark[1,†]

[1]*Department of Physics and Soft Materials Research Center,*
*University of Colorado, Boulder, CO 80309-0390, USA*

[2]*Advanced Light Source, Lawrence Berkeley National Laboratory, Berkeley, California 94720, USA*

[3]*Department of Chemistry and Biochemistry and Soft Materials Research Center,*
*University of Colorado, Boulder, CO 80309-0215, USA*

[4]*Department of Materials Science and Engineering, The University of Utah, Salt Lake City, UT, USA*
*and Soft Materials Research Center, University of Colorado, Boulder, CO 80309-0390*

*Abstract*

The twist-bend nematic liquid crystal phase is a three-dimensional fluid in which achiral bent molecules spontaneously form an orientationally ordered macroscopically chiral heliconical winding of molecular scale pitch, in absence of positional ordering. Here we characterize the structure of the ground state of the twist-bend phase of the bent dimer CB7CB and its mixtures with 5CB over a wide range of concentrations and temperatures, showing that the contour length along the molecular direction for a single turn of the helix is approximately equal to $2\pi R_{mol}$, where $R_{mol}$ is the radius of bend curvature of a single all-trans CB7CB molecule. This relation emerges from a model which simply relates the macroscopic characteristics of the helical structure, which is mostly biaxial twist and has little bend, to the bent molecular shape. This connection comes about through the presence in the fluid of self-assembled oligomer-like correlations of interlocking molecules, arising from the nanosegregation of rigid and flexible molecular subcomponents, forming a brickwork tiling of pairs of molecular strands into a duplex double-helical chain.

---

[*] These authors contributed equally to this work.
[†] noel.clark@colorado.edu

*INTRODUCTION*

An important theme in soft materials and liquid crystal (LC) science is to understand the interplay between molecular shape and macroscopic self-organization, and create new motifs of ordering based on the exploration of different molecular shapes. A prime example of such a scenario has been the discovery of the spontaneous formation of chiral ordering in fluids (Pasteur's experiment in fluids), as evidenced by macroscopic chiral conglomerate domains in liquid crystals of achiral bent molecules [1,2]. In these fluid lamellar smectic phases, the confinement of the molecules to layers forces their steric bends to be in a common in-layer direction, giving long ranged polar ordering, and the steric ordering of molecular tilt direction gives macroscopic chirality [3,4]. Diverse motifs of frustrated packing of achiral molecules lead to chiral isotropic liquids [5], chiral fluid three-dimensional (3D) crystals [6], columnar phases [7], helical nanofilament phases [8], and chiral sponge phases [9].

The most recent, and perhaps most exotic, manifestation of spontaneous chirality in a fluid of achiral molecules is the twist-bend (TB) liquid crystal phase, which fills 3D space with a long-range ordered 1D heliconical precession of pure molecular orientation, sketched in Fig. 1c-e. Initially proposed as a theoretical possibility [10–12], the TB phase has been realized [13,14] and intensively studied in a number of systems of bent molecules [15–25], in particular the bent molecular dimer CB7CB, shown in Fig. 1a. This structure is nematic with infinite helical symmetry: there is no coherent modulation of the density accompanying the helix, as evidenced by the absence of non-resonant x-ray diffraction sensitive to positional ordering [15,26], whereas resonant x-ray diffraction, which probes molecular orientation, reveals the helical periodicity [27,28]. A remarkable of feature of this phase is its very short pitch, on the order of four molecular lengths in CB7CB, such as illustrated in Fig. 1d. In typical nematic LC phases, even strongly chiral ones, neighboring molecules differ in orientation by only a few degrees at most. Phases with larger orientational jumps are always accompanied by positional ordering into 1D lamellar, 2D columnar, or 3D crystalline phases. How the TB phase remains a fully 3D liquid in the presence of such strong coherent internal orientational ordering is a key open question.

In order to explore this issue, we have carried out resonant soft x-ray scattering (RSoXS) and birefringence experiments on a series of CB7CB mixtures with nematic liquid crystal 5CB, measuring the helix pitch and cone angle on given samples, which enable ground state helix geometries to be determined for the first time. The results reveal key geometric features of the TB helix and a remarkable inherent geometric relationship between the macroscopic helix and, $B_{mol}$, the nanometer-radius bend curvature of a CB7CB molecule: (**i**) In the TB phase, twist and bend deformations of the orientation field of the molecular long axis (the director field) are much smaller than $B_{mol}$. The structure is principally a twist of biaxial molecular ordering about the long axes, the pitch of which is comparable to $2\pi/B_{mol}$, quantitatively relatable to the molecular bend curvature, in spite of the near absence of macroscopic director bend; (**ii**) The contour length along the molecular long axis direction for a single turn of the helix is the same for all concentrations and temperatures, implying transient association of molecules into oligomer-like chains; (**iii**) The oligomers are helical linear molecular chains of one handedness that combine to form a duplex helical brickwork-like tiled chain of the opposite handedness; (**iv**) The duplex chains behave in turn as helical steric objects which pack to form the 3D phase, thereby hierarchically self-assembling into the heliconical structure [29,30].

In the conventional nematic (N) liquid crystal phase, the ground state is a three dimensional (3D) fluid in which the density is uniform, exhibiting an orientation field of molecular rods in which the local average molecular long axis, the director field, $n(r)$, has uniform orientation in space [31]. The twist-bend nematic LC phase is also a 3D fluid of uniform density, but composed of bent molecules which self-assemble into a chiral heliconical ground state, featuring the helical precession of $n(z)$ in azimuthal orientation $\varphi(z)$ on a cone of angle $\theta_H$ coaxial with an axis $z$ and having a pitch of ~10-nanometer scale (Fig. 1c-e) [26,27,32]. This chiral structure forms by a symmetry-breaking transition from the uniaxial nematic, even though the molecules are structurally achiral. The resulting heliconical director field has bend and twist elastic deformations of $n(z)$ that are nonzero and of uniform magnitude, $B_H$ and $T_H$ respectively, everywhere in space (Fig.1c-f, [14]). Recently, resonant soft x-ray scattering (RSoXS), by virtue of its sensitivity to molecular orientation, has enabled observation of scattering from this TB helix in the bent molecular dimer CB7CB, where the magnitude of the wavevector of the diffraction peak, $|q| = q_H$, is a direct measure of the pitch of the TB helix, $p_H = 2\pi/q_H$ [27,28]. Combination of $q_H$ data with measurements of the average heliconical cone angle $\theta_H$, suffices to determine the essential mean geometry of the TB helix, including its magnitude of director bend $B_H$.

*RESULTS AND DISCUSSION*

*Helix Pitch and Cone Angle Data* – The phase behavior of binary mixtures of CB7CB and 5CB was characterized as a function of weight percent 5CB, $x$, in the range $12.5 < x < 95$, generating the phase diagram of Fig. 2a (using differential scanning calorimetry, shown in Supplementary Fig. S1). The N–TB phase transition was probed using polarized transmission optical microscopy (PTOM) on the various mixtures in untreated and planar cells for $x$ up to 62.5 (Supplementary Fig. S2). In the mixtures with $x \leq 25$, the TB phase exhibits the typical stripe texture on cooling, which was shown to be due to a spontaneous undulation with displacement along $z$ of the planes of constant azimuthal orientation $\varphi$, [33], caused by dilative stress on the helical structure due to the shrinking of $p_H$ on cooling [27]. For $x \geq 37.5$, we found that the stripes form briefly on cooling, with the texture relaxing into a state of uniform birefringence indicating that the addition of 5CB fluidizes the phase so that it can anneal into an undulation-free director field configuration [34] (See Supplementary Fig. S3). This demonstrates the utility of making mixtures of 5CB with CB7CB to obtain a uniform and undulation-free, well-aligned TB phase. The isotropic (I) to nematic (N) transition temperature of the mixtures, $T_{IN}$, decreases monotonically as $x$ increases, while the N–TB phase transition temperature, $T_{NTB}$, decreases nearly linearly with the addition of 5CB until it becomes undetectable for $x > 62.5$ (Supplementary Figs. S1 and S4). These behaviors are similar to those in mixtures of CB9CB/5CB [19] and CB7CB/7OCB [35].

Resonant soft x-ray scattering (RSoXS) was used to provide a direct *in situ* measurement of the bulk heliconical nematic structure in the mixtures [27], exhibiting diffraction arcs from periodicities in the 8–12 nm range in the TB phase. Extensive freeze fracture study (see Supplementary Information), shows the helix pitch $p_H$, the distance along $z$ for a $2\pi$ circuit of the helix, to also be in the 8–12 nm range [26]. This, and the discussion below, leads us to assign the RSoXS diffraction to be from the fundamental periodicity of the helix, *i.e.*, with the $q_H = 2\pi/p_H$. In the mixtures, the diffraction arcs are smooth and have relatively narrow width in wavevector $q$ at low temperature, as in CB7CB [27], indicating that the pitch is homogenous throughout the sample. As the temperature is raised the scattering tends to broaden in $q$, into a distribution of individual arcs of differing $q_H$, including some with linewidths comparable to 0.0002 Å$^{-1}$, the wavevector resolution of the diffractometer (Supplementary Figs. S5, S6, and [27]). This behavior was taken as evidence for the development of domains in the sample having a distribution of values of average heliconical pitch. Studies in detail in neat CB7CB [27], showed that the lower limit of the pitch in the distributions was a repeatable function of $T$ but that the upper limit varied erratically from scan to scan in $T$, indicating that the TB helix pitch is much softer in response to stretching than to compression, and that the stretching is due to nonuniform stress distributions that develop in the macroscopic textures of the helix axis [27]. For this reason, $p_H$ is taken to be the that from the lowest-$q$ half-height value in each of the distributions, as shown in Supplementary Fig. S6. The resulting $p_H$ data are plotted vs. $T_{NTB} - T$ in Fig. 2b and vs. $T$ in Supplementary Fig. S7. If the temperature of the TB phase of the $x = 0$ and 12.5 mixtures is increased toward $T_{NTB}$, a coexistence range of $T$ is entered in which some of the TB domains melt. This causes the $p_H$ distribution to narrow at the highest $T$'s in the TB phase, the upper limits approaching the lower. However, the $x = 25$ and 37.5 mixtures do not exhibit this behavior: the pitch range remains large near the transition, indicating that the coexistence range is narrower at higher 5CB concentrations, which is also what we observe by PTOM.

The collective optical cone angle of the TB helix, $\theta_H$, was determined for the mixtures from measurements of the birefringence, $\Delta n$, of the N and TB phases, as detailed in the Supplementary Information (Supplementary Fig. S8) [23]. Published values of $\theta_H$ from birefringence [36] and NMR experiments [37] are also available for neat CB7CB. The $\theta_H$ data are plotted vs. $T_{NTB} - T$ in Fig. 2c and in

Supplementary Fig. S9. The data show that $\Delta n$ increases continuously with decreasing $T$ in the N phase, and then near the N–TB transition abruptly begins to decrease. We take this change to indicate the onset of the collective heliconical ordering in the TB phase. The N phase birefringence is somewhat smaller than that of pure 5CB, which we take to indicate the degree of tilt of the cyanobiphenyl groups in CB7CB away from the average orientation in the N phase, which is substantial: ~ 30º (Supplementary Table S1). The heliconical ordering then further reduces $\Delta n$ in the TB phase.

*Bend Deformation and the Geometry of the TB Helix* – If the TB phase heliconical ground state axis is taken to be along $z$, then $\mathbf{n}(\mathbf{r})$, may be written as $\mathbf{n}(z) = (\mathbf{x}\sin\theta_H\cos\varphi(z) + \mathbf{y}\sin\theta_H\sin\varphi(z) + \mathbf{z}\cos\theta_H)$, where $\varphi(z)$ is the azimuthal angle, given by $\varphi(z) = q_H \cdot z = (2\pi/p_H)z$, as sketched in Figs. 1c-f. The heliconical structure can be represented as rotation on a cone as in Fig. 1c or by the green director contour lines in Figs. 1e and f, representing the path (contour line) along which the incremental displacement is always along $\mathbf{n}(z)$. The local nematic order tensor is biaxial, with principal axes given by the director ($\mathbf{n}$), polarization ($\mathbf{p}$), and auxiliary ($\mathbf{m}$) unit vectors. We begin by considering the bend deformation of $\mathbf{n}$, given generally by the director rotation vector $\mathbf{B}_n(\mathbf{r}) = \mathbf{n}(\mathbf{r}) \times [\partial \mathbf{n}(\mathbf{r})/\partial s]$, where $s$ is the displacement along the contour. Since by definition of the pitch we have $q_H \equiv d\varphi/dz$, and from the geometry of the helix in Fig. 1f, $\cos\theta_H = dz/ds$, we find $d\varphi/ds = q_H\cos\theta_H$. According to Fig. 1f, then, the magnitude of the bend of $\mathbf{n}(z)$ in the helix (H) is $B_H = B_n(\theta_H) = (q_H\cos\theta_H)\sin\theta_H$. $B_H$ values calculated using this result are plotted in Fig. 3 vs. $\sin\theta_H$ for the $q_H$ and $\theta_H$ data of Fig. 2. The $B_H$ values fall quite closely onto a straight line passing through the ($B_H$, $\sin\theta_H$) origin, indicating that changing $T$ or 5CB concentration $x$ just moves points along the line, a quite surprising result that enables an immediate prediction: if $B_H(\sin\theta_H)$ is indeed linear in $\sin\theta_H$, then we must have $q_H(\theta_H)\cos\theta_H = S$, the (constant) slope of "the line". At high temperatures in the TB phase where $\theta_H$ is approaching zero, we will have $\cos\theta_H \approx 1$ and therefore $q_H = S$. Thus, for data on the line, the slope of the line should give the limiting helix pitch near the N–TB transition as $p_{Hlim} = 2\pi/S$. Fitting the $B_H$ data in Fig. 3 to a line through the origin yields $S = 0.64$ nm$^{-1}$ and therefore $p_{Hlim} = 9.8$ nm. This value is plotted as the yellow dot in Fig. 2b, and is indeed close to the measured pitches at high $T$ in the TB phase, deviating from the maximum pitches of the different mixtures by less than 10%, characteristic of the deviation of the $B_H(\theta_H)$ data from the fitted line, which has $p_H$ increasing weakly with increasing 5CB concentration.

The observation that $q_H\cos\theta_H$ is nearly constant leads immediately to the question of how to interpret this fitted value of $S$. That $B_H = S\sin\theta_H$ means that $S = B(90º)$, the maximum achievable value of bend of $\mathbf{n}(z)$, obtained when $\theta$ is extrapolated to 90º. However, in the helix we also have director twist, $T_H(\theta_H) = q_H\sin^2\theta_H$, which, if $q_H\cos\theta_H = S$, is given by $T_H(\theta_H) = S\sin\theta_H\tan\theta_H$ for the magnitude of the twist deformation, meaning that the twist elastic energy density $U_T = K_T T_H^2/2$ grows strongly with increasing $\theta_H$, effectively setting an upper limit of $\theta_{Hmax} \lesssim 35°$ on the actually achievable range of $\theta_H$, and making the limit $\theta_H \to 90°$ nonphysical for the TB helix.

*Pure Bend Regime* – We propose alternatively that the extrapolation of $B(\theta)$ to $\theta = 90°$ represents a completely different physical situation, the one exhibiting the maximum preferred bend of $\mathbf{n}(z)$. This must require the geometry: (**i**) in which there is only director bend (pure bend (PB)), (**ii**) in which this bend has its preferred value everywhere (constant magnitude of director bend), and (**iii**) in which $\theta = 90°$, that is $\mathbf{n} \perp z$. These conditions are uniquely realized in the geometry of Fig. 3c, in the system of CB7CB molecules in which their atoms are attracted to a cylindrical surface of variable radius, packed, and equilibrated. At low temperature, this condition maximizes the number density, a condition explored by packing DFT-based (DFT/B3LYP/6-31G**) space-filling models of rigid all-trans CB7CB. Maximum density in a geometry of pure bend is achieved when the molecules are arranged with $\mathbf{B}(\mathbf{r})$ and $\mathbf{n}(\mathbf{r})$ parallel to the *x-y* plane ($\theta = 90°$), and on the cylinder of preferred radius, $1/B_{mol}$. Since the shape of an extended CB7CB molecule matches a circle reasonably well the preferred PB radius can be estimated from the construction shown in Fig. 3b, which minimizes mean square atomic distance from a circle by varying the circle radius $R_{mol}$. That is to say for CB7CB we take the preferred bend $B_{mol}$ in Fig. 3c to be the inverse of $R_{mol}$ from Fig. 3b. The resulting effective molecular radius of curvature of CB7CB is found to be $R_{mol} = 1.58$ nm. This corresponds to a molecular bend of $B = 1/R_{mol} = 0.63$ nm$^{-1}$, which is remarkably close to the slope $S = 0.64$ nm, independently derived in Fig. 3a from the $\theta_H$, $p_H$ data. These data are plotted in Fig. 3a, with $S$ as the red half dot and $B_{mol}$ as the yellow half dot at $\theta = 90°$, where $B(\theta)$ extrapolates to $S$.

We thus find, remarkably, that the $B_H$ vs. $\sin\theta_H$ trajectory of the data in Fig. 3 apparently could have been predicted from the molecular quantity $R_{mol}$, even at small $\theta_H$ where there is little director bend left in the structure, and, furthermore, that the limiting pitch at the N–TB transition can be given in terms of the molecular quantities as $p_{Hlim} = 2\pi/S = 2\pi/B_{mol} = 2\pi R_{mol} = 9.8$ nm. That is, if the TB system $B_H$ vs. $\sin\theta_H$ trajectory has slope $B_{mol}$, meaning that $q_H\cos\theta_H = B_{mol}$, then the helix pitch at $\theta_H = 0°$ is just the circumference of the circle in Figs. 3b and 3c, describing the molecular radius of bend curvature, $R_{mol}$. This circumference accommodates about four CB7CB molecules of all-trans length (~2.6 nm) with a slight overlap of the CN groups (Fig. 3c).

Thus, if the rationale for the twist-bend phenomenon is based on effects of molecular bend, then this observation suggests that the $q_H(\theta_H)$ data respond to changes of temperature and concentration by moving on the trajectory $B_H \approx S\sin\theta_H$, which is in turn being controlled through $S$ by molecular bend by way of $R_{mol}$, even at small $\theta_H$. This result is also surprising because, for TB phases and especially near the N–TB transition where $\theta_H$ tends to be the smallest, there is little director bend left in the structure, the bend magnitude $B_H(\theta_H)$ getting quite small, as shown in Fig. 3. Nevertheless, if the data are on the line, bend is still proportional to $B_{mol}$. Thus, the data of Fig. 3 indicate that the structural preference for the TB ordering to give a well-defined $q_H$, is definitely not a preference for constant bend. Several studies analyzing the elasticity of the TB helix pitch have recently found the director curvature bend energy to be orders of magnitude too weak to account for the TB pitch compressional elastic constant, $C$, measured as $\theta_H$ becomes small, and were led to propose local lamellar smectic positional correlations as an alternate source of rigidity [21,38]. Fig. 3 shows, however, since the data are nearly on the line, that the helix pitch appears to be controlled by molecular bend, even in the absence of director bend at small $\theta_H$.

*Polygon Chain Model* – These considerations lead next to the question of the geometrical meaning of the line and, in particular, the relationship $d\varphi/ds = q_H\cos\theta_H = B_{mol}$. Thus, on the line, for different values of $\theta_H$, $q_H$ is such that $s(\Delta\varphi = 2\pi)$, the net distance traveled along the director contour path for a $2\pi$ increase in $\varphi$ (one complete turn of the helix, shown as the black dashed line in Fig. 1e), is independent of $\theta_H$, and furthermore given by $s(\Delta\varphi = 2\pi) = 2\pi/B_{mol} = 2\pi R_{mol}$ for all $\theta_H$, including the PB regime $\theta = 90°$ (the path through $\Delta\varphi = 2\pi$ around the circle of circumference $2\pi R_{mol}$). The equality $s(\Delta\varphi = 2\pi) = 2\pi R_{mol}$ does not appear to be a symmetry of the system, because the TB ground state at small $\theta_H$ is entirely different from that of the PB at $\theta = 90°$. However, this is clearly this condition that connects the two regimes.

We can shed light on this condition by developing a geometrical model, the rectangle-triangle (RT) polygon chain, sketched in Fig. 4, that, by design, exhibits $s(\Delta\varphi = 2\pi) = 2\pi R_{mol}$ over the entire range of $\theta_H$. This model is an assembly of rectangular and triangular plates, connected into a periodic chain where the lines representing shared edges of rectangles and triangles are bendable hinges. The rectangles are attached to, and constrained to be locally parallel to, a flexible rod in the form of a helical spring representing the contour line of the director (green lines, Figs. 3e, 3f, 4b), on a cylinder of tunable radius $R = R_{mol}\sin\theta_H$. The corresponding tunable pitch $p_H = 2\pi R_{mol}/\cos\theta_H$ guarantees that such a chain of length $s(\Delta\varphi = 2\pi) = 2\pi R_{mol}$, in this case this length of its eight rectangles, $8L$, always makes exactly one turn of the helix. The rod is inextensible, enforcing by construction the condition that $s(\Delta\varphi = 2\pi) = 2\pi R_{mol}$. In the basic structural unit of the chain, consisting of two rectangles and a 45° isosceles triangle, a 45° bend in the director is enforced by the triangular hinge when the triangle and rectangles are all in the same plane, the condition at $\theta = 90°$ where the whole construction lies parallel to the *x-y* plane (Fig. 4a). This directly models the molecular organization of the PB limit in the cylindrical shell packing of Fig. 3c.

The hinge angle $\beta_o = 45° = 360°/8$ was chosen because, as discussed below, a diffuse feature in the non-resonant x-ray scattering indicates that that there are ~8 molecular half-lengths in the TB pitch at small $\theta_H$. Since the data are on the line there must correspondingly be in the PB regime ~8 segments around the $2\pi R_{mol}$ circumference, and indeed, as shown in Fig. 5a, the PB regime is well modeled by the arrangement of four 45° bent rod molecules. We propose that each ring in this structure is stabilized by neighboring rings, in an arrangement where adjacent rings have a difference in azimuthal orientation of 45°, such that the flexible molecular centers in one ring are over the regions of fluctuating end-to-end molecular contacts in the neighboring ring, an entropically favored association. This makes a construction like a cylindrical brickwork chimney, discussed in the next section. With this choice, eight rectangle long edges must make a complete turn, so the rectangle length, $L$, is chosen such that $8L \sim 2\pi R_{mol}$. The corresponding magnitude of director bend is then $B = (\pi/4)/L = (2\pi)/8L = 1/R_{mol} = B_{mol}$.

The helix can be tuned by pulling the ends of the rod so that they become separated along a line parallel to $z$, the separation being the pitch $p_H$, as indicated by the black arrow in Fig. 1e, which decreases $\theta_H$ and makes the rod less bent everywhere along its length. The bend angle $\beta$ of the local elements thus decreases from the maximum of $\beta_o = 45°$ causing them to buckle, the triangle swinging out of the plane of the rectangles to affect less bend, and in the process inducing a local relative twist $\tau$ of the rectangle planes, which are free to rotate about the rod axis, as sketched in Fig. 4a. With this geometry, if the tilt of $n(z)$ relative to $z$ is $\theta_H$, then the angle between the triangle and rectangle planes will be $-\theta_H$, the condition which keeps the triangle planes always parallel to the $x$-$y$ plane. If the separation of the rod ends is increased and $p_H$ approaches the rod length $2\pi R_{mol}$, then $\theta_H \to 0$, $B_H \to 0$ as $B_H = \sin\theta_H/R_{mol}$, and the rod becomes nearly straight, with the local geometry in Fig. 4a. The triangle plane is now normal to the rod, and its initial induced bend in the rod of $\Delta\varphi = 45°$ is now completely converted to an induced local relative twist about the director of the rectangle plane normals through $\Delta\varphi = 45°$ at each hinge, as in Fig. 4b. The bend angle, $\beta$, twist angle $\tau$, and $\beta_{mol}$ are geometrically related as shown in Fig. 4a.

The rectangles also represent the principal axes of the local biaxial nematic ordering tensor of the director field (director $n$, flexoelectric polarization direction $p$, auxiliary unit vector $m$), as in Fig. 4a. Thus, as $\theta_H$ increases the overall structure of a single pitch is converted from the $\theta_H \sim 90°$ state: a series of 8 steps of 45° rotation of director bend and of local biaxiality about its $m$ axis on the circumference of the circle of radius $R_{mol}$; to the $\theta_H \sim 0°$ state: a series of 8 steps of 45° twist rotation of the local biaxiality about its $n$ axis, on a path along $z$ of length $2\pi R_{mol}$. This scenario precisely maintains $q_H\cos\theta_H = B_H$ throughout the range of $q_H$, i.e., puts $B_H$ vs. $\sin\theta_H$ on the line. We denote these ranges of large and small $\theta_H$ respectively as the pure bend (PB) regime ($\theta_H \sim 90°$) and the twisted biaxiality (TBX) regime. The RT model shows directly that the structural stability of the local elements through the transition from pure bend to twisted biaxiality is what is required to maintain the compressional elasticity of the pitch under the condition that $B_H \to 0$ and director curvature elasticity drops out. Actual twist-bend phases typically have $\theta_H \lesssim 30°$, so they are much closer to the twisted biaxiality limit than the pure bend. Thus, in "TB" the twist should be taken to mean twist of the biaxiality. The $\theta_H \sim 0°$ regime represents the state of the helix dominated by twisted biaxiality but having no macroscopic optical tilt. Such a state is achievable as will be shown below. In the CB7CB mixtures the TB phase appears to come in with a small but finite $\theta_H$, consistent with the optical, x-ray and DSC evidence for a first order N–TB transition.

The RT model can be made for any angle $\beta_{mol}$. If $\beta_{mol}$ is small, then $\tau^2 + \beta^2 \cong \beta_{mol}^2$, with $\beta$ and $\tau$ becoming the orthogonal projections of a vector of magnitude $\beta_{mol}$, constrained to move on a circle (Fig. 4a). In the limit that $\beta_{mol} \to 0$ with $L/\beta_{mol}$ constant and assuming that the hinge bends remain highly flexible, the RT chain becomes a like a sheet of paper bent into the accordion fold of a fan, with high bending rigidity in the radial direction, and low bending rigidity in the circumferential direction. Upon pulling the bend out, such a sheet will exhibit little elastic resistance against conversion from continuous bend to continuous twist. In the continuum limit $B_H^2 + T_{BX}^2 = B_{mol}^2 = B_H^2 = \beta_{mol}^2/L^2$, the result also derived in Fig. 1f from the projective geometry of the helix. The exchange of bend and twist is controlled by $B_{mol}$ even in the limit of zero bend ($T_{BX} \cong B_{mol} - B_H^2/2B_{mol}$).

*Steric Oligomerization of Bent Molecular Dimers* – In the RT model, the constraint that $q_H \cos\theta_H = B_{mol}$, independent of $\theta_H$, is built into the model by the fixed length of its chain of polygons, a condition that would seem most applicable to a system of locally-bent flexible oligomer or polymer chains. In the dimer TB phases considered here, there are no chemical links between molecules, so it is necessary to understand, in the context of independent bent molecules, how such a similar polymer-like condition could come about in both the PB and TBX regimes, how the PB and TBX regimes are linked, and, therefore, how biaxial twist in absence of bend comes to be controlled by $B_{mol}$.

We propose that molecular bend and steric packing constraints of the condensed TB phase combine to stabilize oligomeric chains of molecules, and that the brickwork packing motif, introduced in Fig. 3c and detailed in Fig 5a, is the common structural feature that stabilizes the chains and connects the PB and TBX regimes. The brickwork packing of a pair of adjacent chains can be visualized as a string of segments, each containing a pair of oppositely directed molecular halves linked by interfaces, each containing the center of a molecule in one chain and the tails of two in the other. This motif has also been found in molecular dimer liquid crystal structures [39]. This assembly is stabilized by the well-known tendency for rigid and flexible molecular subgroups to nano-phase segregate [40], with the flexible molecular centers most readily accommodating the fluctuations in relative position of neighboring molecular ends or tails. We refer to a double helix chain formed in this way as a duplex helical tiled chain (DHT chain, DHTC). The intraduplex tiled linking is responsible for the apparent fixed contour length, $s(\Delta\varphi = 2\pi) = 2\pi/B_{mol}$ along $\mathbf{n}(\mathbf{r})$ and manifested in the construction of Fig. 4.

This proposal is supported by the observation of a diffuse non-resonant x-ray scattering feature in the N and TB phases of pure CB7CB, having a peak on the $q_z$-axis, at $q_m \approx 5.05$ nm$^{-1}$ [15]. A similar peak is found in the N and/or TB phases of a variety of other bent molecular dimers, with $q_m$ in the range 4 nm$^{-1}$ < $q_m$ < 5 nm$^{-1}$ [13,15,19,22,23,28,41,42], as discussed in Supplementary Figs. S11-S13. The typical appearance of this TB phase feature is shown in Fig. S11, which plots the non-resonant x-ray structure factor of the TB phase in CB7CB, calculated from the molecular dynamics simulation in the TBX regime reported previously [26]. The white ellipses indicate the on-axis peaks, which can also be seen in Supplementary Fig. S13, which plots $z$-axis intensity scans $I(q_z)$ of CB7CB [15] and of a DTC5C7 /DTSe mixture [28]. These scans indicate a periodic electron density modulation and therefore molecular positional ordering along the helix with a fundamental periodicity of $d_m \cong 1.25$ nm, consistent with the presence of short ranged periodic positional correlation of similarly structured molecular segments along $z$. This finding supports the brickwork association proposal since this value of $d_m$ is close to half of the molecular length $M = 1.4$ nm of extended CB7CB, which is that required for the segment length in a brickwork tiling. In fact, comparison of $d_m$ with extended molecular length for the bent molecular dimer systems for which $d_m$ data is available, noted above, shows that the condition $d_m/M \sim 1/2$ appears to be a general trend, as illustrated in Supplementary Fig. S12. In CB7CB, since $d_m$ is comparable to $p_H/8 = 1.22$ nm, it is close to the brickwork segment length in the PB regime on the $R_{mol}$ circle (Fig. 5a), an observation that can be taken as evidence for there being similar segments at small $\theta_H$.

We made an initial evaluation to explore whether the peak at $q_m$ (the chain segment scattering) could be understood on the basis of a model in which the self-assembly of a pair of molecular chains is described as a periodic chain of half molecule-long segments, each connected to an adjacent segment by a nearest neighbor harmonic spring. The structure factor of this standard model for 1D positional ordering exhibits only short range order at finite temperature [43,44], as described by the monotonic increase of the mean square of fluctuations in the dynamic separation of pairs of elements of the chain with increasing mean separation: $<(u_0 - u_n)^2> = <\delta u_n^2> = \sigma^2 n$, where $n$ counts the segments along the chain. The 1D structure factor $I_{1D}(q_z)$ fits the data on CB7CB [15] and the DTC5C7/DTSe mixtures [28] quite well (Supplementary Fig. S13), giving, in both cases, a distance along the chain of ~15 segments for translational order to be lost, i.e., where $\sqrt{<\delta u_n^2>}$ becomes equal to the segment interval $d_m$.

Another common feature of the chain segment scattering, $I(q_z,q_\perp)$, in the materials in Supplementary Fig. S12 is that the width of the diffuse peak in the direction normal to the helix axis $z$, $\delta q_\perp$, is significantly larger than $\delta q_z$, its extent along the helix, as in Supplementary FIG. S11. In some cases this appears to be due to mosaic broadening due to alignment defects, but in the TB phase in the DTC5C7/DTSe mixtures [28], for example, the narrow angular width in $q$ of the resonant $q_H$ peaks shows that the sample is well aligned, and therefore that the broadening of $I(q_z,q_\perp)$ in the $q_\perp$ direction is intrinsic. The corresponding correlation lengths $\xi_\perp$ and $\xi_z$ have the inverse relationship, implying that the correlation volumes giving the diffuse non-resonant scattering are extended along $z$, *i.e.*, a chain-like periodicity along $z$ rather than layer-like correlations.

*Duplex Helical Tiled Chain (DHTC) Structure of the TB Phase* – The challenge then is to develop a model of the TB helix with small $\theta_H$ in which it is made up of at least pairs of molecular chains in a brickwork tiling with subsections along $z$ of pairs of antiparallel half-molecules, in which, for CB7CB the structural twist between segments is 45º. To this end, we considered the organization of single stick and space-filling molecular models consistent with the above requirements. The PB regime is readily modeled by the packing of all-trans space-filling models of CB7CB and, as in Fig. 5, by either two or three segment bent stick models having 45º or 30º bends, respectively. in the PB limit, brickwork tiling of either stick model gives a bend of 45º per segment and four molecules per ring (Fig. 5a), so that the change in azimuthal orientation $\varphi$ is 45º per segment.

The required structures are shown in various representations in Figs. 5b,c and 6, and Supplementary Fig. S16. The basic structural associations are of three molecules like that of the green, cyan, and yellow groups inside the black elliptical rings in Figs. 5a,b, wherein terminal groups of the cyan and yellow tuck into the volume of hard-to-fill space vacated by the bending of the green, and can associate with the flexible central aliphatic linkers. This scenario is repeated for the next segment along $z$, among a group rotated through 45º relative to the initial one and having the cyan molecule in the center, and so on for all $z$. The stick models in Figs. 5b,c and 6 show that this structure is double-helixed, made up of two identical right-handed helical chains of molecules, each transforming into the other by a translation of a single segment length followed by a 45º rotation in azimuthal angle. The paired assembly of two chains is stabilized in both the PB and TBX cases by a combination of a constraint, in the former to be on the cylinder or in the latter to be in a tube created by neighboring chains, and by the pressure exerted by the neighboring molecules. In the pairing of the single strand chains, the overlaps stabilize the structure and the interlocking bends promote the filling of space. In the bent stick representation of Fig. 5c, the half molecular rods can be taken to represent the optical anisotropy of the halves of the molecule. Taking this half molecule polarizability to be uniaxial, the effective optical anisotropy of a segment of the double helix can be obtained using the construction in Fig. 5c. Here, the white square at each level is marked with a black dot that marks the midpoint between the intersections of the two chains with the square. The dark green line connects the midpoints from square to square. Thus, in a given segment the green line construction will give the orientation of the local principal axis of the average dielectric tensor having the largest refractive index, which we take to be the local director. Thus, the green line represents the trajectory of the optical $n(r)$, which is also a right-handed helix. This construction shows that, in a given segment of the DHT chain, the tilts of the half-molecular optic axes away from $z$, in this case by ~ 22º, tend to cancel one another, leaving a much smaller effective optical heliconical cone angle, in this case $\theta_H$ ~ 11º. The magenta labels in Fig. 5c indicate the handedness of the various helices, with the single chains and the director helix being right-handed (RH). Interestingly, the double helix is left handed (LH).

For clarity, the molecules in Fig. 5 are positioned with more symmetry than they will actually have in the typical case. Generally the planes of the bent molecules in the helices of Fig. 5b,c will be tilted away from $z$, through an angle $\psi$ as shown in Fig. 6 and in Supplementary Fig. S16 for both signs of tilt from $z$. The untilted case could occur at some particular temperature, like the unwinding of the helix in a chiral nematic at a particular temperature.

Fig. 6 presents the fully formed DHTC structure in the pure TBX limit for which the optical director tilt $\theta_H = 0º$. In this structure, the projection of the halves of a given tilted molecule onto the x-y plane has an opening angle between them of 45º (Fig. 6c), the same as the rotation $\Delta\varphi = 45º$ per segment. This condition requires a tilt of the molecular plane from $z$ of 9.9º. In this case, the two molecular halves in a given segment have parallel projections onto the x-y plane (Figs. 6a,c). Since they also have equal and opposite tilts there must be a principal axis along $z$ of their average biaxial contribution to the dielectric tensor (Fig. 6d). Starting from this structure a heliconical director field of finite $\theta_H$ can be generated by changing the molecular tilt (Supplementary Fig. S16) or by helical deformation the DHT chain (Supplementary Fig. S16). Introducing director bend into the DHT chain reduces biaxial twist, following the geometric projection scenario of Fig. 3f and of the RT model in Fig. 4. This comes about as illustrated in Figs. 6b,e, showing that, on the boundary between the two duplex chain segments containing the halves of the red molecule (denoted by a black circle), the projections of the halves of its cyan and yellow molecular neighbors make a 45º angle to one another. As indicated in Fig. 6b, this corresponds to twist $\tau = 45º$ for $\beta = 0$ at a yellow triangle in Fig. 4a. In the presence of director (heavy green line) bend, $\beta$, the rotation of these neighbors relative to the red molecule is of opposite sign ($+\beta/2, -\beta/2$) and applied on the projections, as on the edges of the yellow triangle, causing the black disc plane, with application of bend, to rotate about $p$, remaining, as in Fig. 6b, coplanar with the yellow triangle as it reorients (Fig. 4a). Elastic deformation of the DHT chain then satisfies the RT model constraints which put $B_H(\theta_H)$ on the line in Fig. 3a: $q_H\cos\theta_H = B_{mol}$, $B_H = B_{mol}\sin\theta_H$, $T_{BX} = B_{mol}\cos\theta_H$, and $B_H^2 + T_{BX}^2 = B_{mol}^2$. For $B_H$ small, then, the reduction in biaxial twist is controlled by $B_{mol}$, with $T_{BX} \approx B_{mol}(1 - B_H^2/2B_{mol})$. In the limit of large bend twist is eliminated and the structure evolves toward the PB limit (Fig. 6e).

*Three-Dimensional Heliconical State* – The bulk TB phase is a 3D space-filling packing of DHTCs. The overall orientational ordering with uniaxial positive birefringence means that the DHT chains are generally running parallel to one another, making the TB a hierarchical nematic self-assembly of anisotropic self-assembled oligomeric chains. In the packing of cylindrical objects that are helically modulated, the helical contours on adjacent facing cylinders cross each other (like the stripes on a pair of parallel barber poles of the same handedness if put into contact). This geometry tends to suppress melding of the chains and to maintain the cylinders as distinct entities in the packing. Each DHT chain then is effectively confined to an on-average cylindrical hole in the fluid by its neighboring chains which exert an effective pressure like that coming from osmotic pressure in a depletion interaction. This picture is supported by the experimental finding that the $B_H(\theta_H)$ data of all the mixtures lie on the same line in Fig. 3a, indicating that they behave as if they all have the same $B_{mol}$ (at $x = 37.5$ we might have expected a significant dilution effect leading to a smaller $B_{mol}$). The constancy of $B_{mol}$ suggests that in the structures determining the pitch, the DHT chains in the case of in the TBX, are comprised dominantly of the bent dimers, and that the 5CB is a filler in between. The 5CB dilution lowers the phase stability and reduces $\theta_H$, but this all occurs with $q_H \cos\theta_H = B_{mol}$, implying chains under the same constraint: $d\varphi/ds = B_{mol}$.

Next, we consider the steric packing of the DHT chains that makes up the bulk phase. The fact that the resonant x-ray scattering from the bulk TB exhibits diffraction spots from oriented domains that are 3D smectic-like, that is, having resolution limited width in $\delta q_\perp$, indicates that the long range ordered pseudo-layer scattering objects are arrays of lamellar sheets extended in the in-plane direction [27,28]. This means that in the bulk TBX packing, the phase $\varphi$ of the twist in a DHT chain must become coherent with that of its neighbors, a condition that has been observed in nematic phases made by packings of chiral particles internally structured as a steric repulsive helical line, realized, for example in suspensions of helical flagellae [29], and in the extensive simulations of steric helices of Kolli *et al*. [30]. Another example relevant to the TB phase is the helical nanofilament phase found in neat bent-core systems [8,45] in which chiral filamentous bundles of a few smectic layers achieve macroscopic phase coherence of their twist solely by interacting through their periodic biaxiality.

The Kolli simulations appear to be particularly applicable to describe the interaction of, and the potential of long range phase ordering for, the DHT chains for finite $\theta_H$ in the TBX regime. Supplementary Fig. S14 shows an example of the systems of interacting particles employed by Kolli *et al*., composed of rigid helical chains of contour length, *L*, made of truncated hard spheres of diameter, *D*. Comparison of the Kolli particles with the steric shape of the DHT chains of CB7CB, made in Supplementary Fig. 12 for the $\theta_H = 10º$ case in Fig. 5b shows particles with helical radius $r/D \approx 0.2$ and pitch, $p \approx L \approx 10D$ match the CB7CB DHT chains quite well. The Kolli *et al.* phase diagram for $r/D = 0.2$, also reproduced in Supplementary Fig. S14, shows that particles having $r/D = 0.2$ systematically give I, N, TB and smectic phases, with the TB range decreasing as the pitch becomes comparable to and longer than the particle contour length. Thus, the single pitch duplex CB7CB chains should be able to order into a 3D TB phase if sufficiently long and rigid. The TB range in Supplementary Fig. S14 is limited with increasing volume fraction by the appearance of smectic ordering, corresponding to the positional ordering of the particles into smectic layers of thickness comparable to their length. In the case of living polymer chains, like what we propose for CB7CB, the effective particles will be transient and polydisperse, the latter condition well known to strongly suppress smectic ordering [46], an effect which may expand the TB range.

For $p \approx L$ and $r/D = 0.2$ the Kolli *et al.* helical particles behave as if they are smooth, like those of Barry *et al.* [29], which have helical glide symmetry. In these cases, if the steric helical interaction is reduced, for example, by reducing $r/D$ or making *pitch/L* large, the system will revert to a simple nematic or smectic phase. However, the DHTCs are not smooth, but are periodically structured, with a local biaxial shape, so the role of variations of the steric shape along the DHT chain must also be considered. Figs. 4b, 5c, 6, S11, S14, and S16 all exhibit aspects of the biaxiality of the DHT chains. Fig. 4b for the $\theta_H = 0°$ case and Fig. 5c show that the projection of the segments onto the drawing plane varies in effective shape along the chain, with a period equal to half of that of the helix. This variation is also evident in the projection of the steric shape of a duplex chain in Fig. S14. Generally, each segment is biaxial, with a steric cross section in the x-y plane that has the symmetry of an ellipse. This elliptical shape rotates in azimuthal angle φ along the chain (biaxial twist), as in a twisted ribbon of zero net local curvature, with a period of $p_H/2$, equal to four segment lengths as is clear in Fig. 4b. In the TBX limit, this is the only periodicity of the DHTC. In a dense packing of the DHTCs steric variations in shape, especially periodic ones will lead to the development of correlations between chain positions along *z*. This will be an especially strong effect if the oligomerization has substantially reduced the translational entropy for displacement along *z*. In the helical nanofilament case [9,45], where the structural periodicity is the helix half pitch and filament steric profile almost circular such that the neighboring filaments only weakly sense each other's grooves, the filaments have a strong tendency to order with their biaxial twist in phase. In the DHTC case, sufficiently large ellipticity and packing density will lead to a 3D structures in which adjacent duplexes will align out of phase to facilitate packing. Twisted ribbons, for example pack best when shifted by a quarter of their pitch.

*RSoXS as a Probe of the Duplex Helical Tiled Chain Model* – Given that we a have now have a fairly detailed structural model, we reconsider RSoXS as a probe of the heliconical structure of the TB phase.

Interestingly, the first applications of resonant scattering to LCs was to probe the heliconical molecular orientational ordering in chiral tilted smectic phases, wherein the molecules are confined to layers, sorting out layer-by-layer sequences of azimuthal orientations of tilted molecules [47,48]. In this context the general theory of resonant scattering was applied to the smectic case [49]. This formalism has recently been applied in a comprehensive analysis of RSoXS scattering from the TB phase by Salamończyk *et al.* [50].

RSoXS at the carbon K edge (incident wavelength, $\lambda$ = 4.4 nm) gives a range of scattering vectors $q < 2\pi/2.2$ nm, probing length scales through the nanometer range down to ~ 2 nm: molecular, but not atomic, size. In this $q$ range, molecular subcomponents such as the biphenyls in CB7CB act nearly as composite entities in the scattering process, being describable by second-rank molecular polarizability tensor scattering cross-sections, as in deGennes' formulation of light scattering by fluctuations in director orientation [31]. In analogy with light microscopy, RSoXS could even be used to visualize patterns of birefringence of LC phases and textures with x-ray resolution using depolarized transmission.

In probing the DHTC model of the TB phase, we first consider the RSoXS from individual filaments with the aid of Supplementary Fig. S15, focusing on the essential qualitative features of the scattering in the simplest geometry. This figure shows about 1.5 pitches of the $\psi = 0$ DHT chain in Fig. 5c, represented by space-filling models of CB7CB. In a typical experiment, the sample cell with the LC between silicon nitride windows is parallel to the Figure plane, the TB helix axes of the LC are aligned parallel to the windows, and we consider illuminating a domain with the local helix axis vertical on the image, as shown. Incident x-rays pass through the image plane and are forward scattered onto a 2D detector behind. The incident and scattered directions can be chosen so that the scattering vector $q$ is parallel to the DHTC $z$ axis. In this example, we take the incident x-ray polarization, $i$, to be horizontal. The helical winding of the filament is apparent in the figure, with the director giving the orientation of a principal axis of the polarizability tensor following a helical trajectory, as in Fig. 1 and, in particular, in Fig. 3c. According to deGennes, the depolarized scattering field amplitude probing director orientation is approximately $E_d(z) \propto (f \cdot z)(\delta n(z) \cdot i)$, where $f$ is the outgoing polarization, nearly parallel to $z$, and $\delta n(z)$ is the angular deviation from $z$. The key feature of this relation is that $E$ is linear in $\delta n(z)$, so that the sinusoidal projection of the helix structure onto the $i$-$z$ plane gives a sinusoidally varying scattering amplitude as $\delta n(z) = \sin(q_H i)$, which in turn produces scattering at $q = q_H$. This is the basis of the claim that the depolarized scattering peak determined the helix pitch $p_H = 2\pi/q_H$, which clearly should be applicable to scattering from a single DHTC. In addition to the helical undulation, the DHTC exhibits smaller scale roughness, a result of the precessing biaxiality discussed in the previous section. Inspection of the DHCT shows that, like in Fig. 4b, there are in general four distinct projections of the biaxial order on any vertical plane. Scattering from these variations has amplitude $E_p(z) \propto (\mathbf{p}(z)\cdot\mathbf{i})(\mathbf{p}(z)\cdot\mathbf{i})$, where $\mathbf{p}(z)$ is the biaxial orientation vector. Here, the scattered amplitude is independent of the sign of $\mathbf{p}(z)$, so that the projection of $\mathbf{p}(z)$ onto a vertical plane $\delta n(z) = \sin(q_H z)$ generates a polarized scattering amplitude $E_p(z) \propto \cos(2q_H z)$, the second harmonic of the scattering from the helix, explicitly showing that the periodicity of the biaxiality is a half-pitch: flipping the rectangles around $\Delta\varphi = \pi$ doesn't change their biaxial polarizability. All the DHT chains presented here share this property. Salamończyk *et al.* have pointed out that in the scattering from columns of helical precessing tilted rods, averaging together pairs of columns shifted relatively in phase by a half period of their biaxial polarizability renders the net polarizability the same in every quarter period and the second harmonic disappears. In the few experiments where the second harmonic might have been seen, it has not been [27,28,50], indicating that such averaging may be taking place in the TB phase. In the case of the DHTCs, shifting a pair of chains by two segments and averaging will eliminate the second harmonic. However, achieving this in arrays of DHTCs may be problematical, since frustration effects come into play on the closest packed 2D hexagonal lattices.

*Asymmetric Elasticity of the TB Helix* – A relevant feature of the lamellar-like helical ordering of the TB phase is its strange asymmetry in the response of the TB helix pitch to compressive or dilative stress [27]. In typical fluid lamellar liquid crystals, such as the smectics A and C, stresses tending to change the layer thickness encountered in typical textures, for example planar aligned cells or in focal conic powders in capillaries, exhibit little observable variation in layer spacing in x-ray scattering experiments, except near phase transitions where the compressional elastic constant can become small. In the TBX regime of CB7CB, the layer spacing exhibits a well-defined minimum trajectory vs. temperature, but it can be significantly increased in irregular fashion from this minimum value by dilative stresses appearing in textures in 1 µm thick sample between flat plates. The DHT chains can be expected to respond in an asymmetric way to stress along $z$, with the end-to-end packing of the molecules in the chains resisting compression, but the steric association of the overall structure being rather soft against stretching, providing a natural explanation for this behavior.

*Model Systems of Bent Rigid Molecules* – We sought to explore the role of molecular bend in other TB systems. The only others of which we are aware and for which data sets of $p_H$ vs. $\theta_H$ are available are the mean-field theoretical model for bend rods of Greco *et al.* [51], and the Monte Carlo simulation of Greco *et al.* of hard spheres assembled to make steric circular arcs [52]. These models are of particular interest because they treat collective TB behavior for bent objects that are rigid. Fig. 7a and b show plots of $B_H$ vs. $\sin\theta_H$ calculated from $p_H(\theta_H)$ for the arcs and bent rods respectively. The black line in each plot is drawn through the origin and $B_H(\theta_H)$ for the smallest $\theta_H$. The general behavior of $B_H(\sin\theta_H)$ is similar to that of the CB7CB mixtures in Fig. 3a, but with a tendency to increase relative to the black line with increasing $\sin\theta_H$, which is also seen weakly in neat CB7CB (Fig. 3a).

In the case of the circular arcs we carried out the $R_{mol}$ construction of Fig. 3b, with the result shown in Fig. 7a. We carried out the construction to determine or the circular arcs, finding $R_{mol} = 12.6$ in units of the sphere diameter, σ. The corresponding $B_{mol} = 1/R_{mol} = 0.08/\sigma$ is comparable to the $B = 0.1/\sigma$ extrapolation of the black line, indicating a relation between the PB and TBX limits similar to that in the CB7CB mixtures. This makes the hard arcs a very interesting system for exploration of the DHTC structure.

Turning to the bent rod case, Fig. 7b shows that $B_H(\theta_H)$ obeys $B_H = S\sin q_H$ rather well for $\theta_H < 15°$, with a slope $S = 0.56/L$, where $L$ is the length of one of the arms in the bent rods. In Fig. 7b we have used the small-angle value of $S$ to extrapolate to $\theta_H = 90°$ in order to determine the radius of the cylinder $R = 1/S$ in the PB regime. The resulting construction using the shape of the simulated bent rods, shows a quite reasonable PB limit. This is an exciting result because this model approaches understanding the TBX regime from a mean-field statistical mechanical approach that is entirely different from the geometrical model building that we have employed. That it captures the essence of the geometry of "the line" offers an opportunity to understand in detail the evolution of the local geometry can keep the system on the line in absence of change in molecular conformation.

Both of these models employ rigid molecules and yet seem to exhibit the same essential geometrical behavior as the CB7CB system, which was rationalized on the basis of nanophase segregation of flexible (the central alkyl linker and the tail ends) and rigid molecular subcomponents (cores): the molecular ends find entropic freedom by associating with the flexible cores. We propose that in systems of rigid bent hard particles the analogous association is between the particle ends and the free volume available in the pocket of difficult-to-fill space created by molecular bend.

*CONCLUSION*

The heliconical structure of the CB7CB mixtures exhibits a remarkable dependence on concentration and temperature in which director bend and biaxial twist exchange under a strict geometrical constraint that equates the magnitude of director bend at one extreme of cone angle to the magnitude of biaxial twist at the other extreme (moving on "the line"). This constancy is related to the ability of the TB phase to maintain inherent bend and a fixed contour length along the director that are both elastically stabilized, behavior most directly understood on the basis of the longitudinal connection (oligomerization) of molecules. This, and the x-ray observation of half-molecular length periodicity along the TB helix leads to the model of the self-assembly of half molecule-long segments into duplex helical tiled chains of molecules as the basic structural element of the TB phase. The geometrical constraint then shows up as $\Delta\varphi = 45º$ director bend jumps from segment to segment in the PB ring limit, and 45º biaxial twist jumps in the TBX limit. As noted in the text there is no symmetry requirement for this equality. The structural organization of the DHT chains should accommodate a range of possible $\Delta\varphi$'s per segment, for example it appears that DTC5C7/DTSe [28] and AZO7 [53], with ratios of $p_H/p_m$ ~6 in the TBX regime are $\Delta\varphi$ ~ 60º/jump, and there seems to be no structural impediment to 90º per jump [39]. Since these $\Delta\varphi$ ~ 60º materials are unlikely to have comparably large bend jumps in the PB regime, they will not move on "the line". Thus, how general the CB7CB behavior is among the TB materials remains to be seen. However, in any case, the CB7CB scenario appears to offer a useful benchmark for relating the molecular structure and macroscopic behavior in TB phases. If the data are not on the line as for the simulated systems in Fig. 7, the deviations can be explored by comparison with CB7CB ideal elastic behavior. For systems exhibiting the CB7CB scenario there remains the question of relating the slope $S$ of the $B_H$ vs. $\sin q_H$ curve to the molecular shape. CB7CB also turns out to be extremely simple in this regard (maybe not just a coincidence). In the case of CB7CB, the extended molecule is almost circular so the $B_{mol}$ is readily determined from the molecular shape and $S$ matches this bend very well. In general, however, it will not be so easy to assess molecular bend, for example, flexibility in molecules with longer aliphatic linkers will likely make $B_{mol}$ smaller than estimates based on extended molecular shape. This leads to the question of how to design calculational or simulation schemes of the pure bend regime that can quantitatively predict the TB structure by determining $B_{mol}$ to get $S$.

*ASSOCIATED CONTENT*

Supplementary Information.

*AUTHOR INFORMATION*

Corresponding Authors

Email: noel.clark@colorado.edu, michael.tuchband@colorado.edu, min.shuai@colorado.edu

*NOTES*

The authors declare no competing financial interest.

*ACKNOWLEDGEMENTS*

This work was supported by NSF MRSEC Grant DMR-1420736, by the Institute for Complex and Adaptive Matter Postdoctoral Fellowship Award OCG5711B, and by ED GAANN Award P200A120014. LR acknowledges support by NSF grant DMR-1001240, and by the Simons Investigator award from the Simons Foundation. We acknowledge use of beamline 11.0.1.2 of the Advanced Light Source supported by the Director of the Office of Science, Office of Basic Energy Sciences, of the U.S. Department of Energy under contract no. DE-AC02-05CH11231.

*FIGURES*

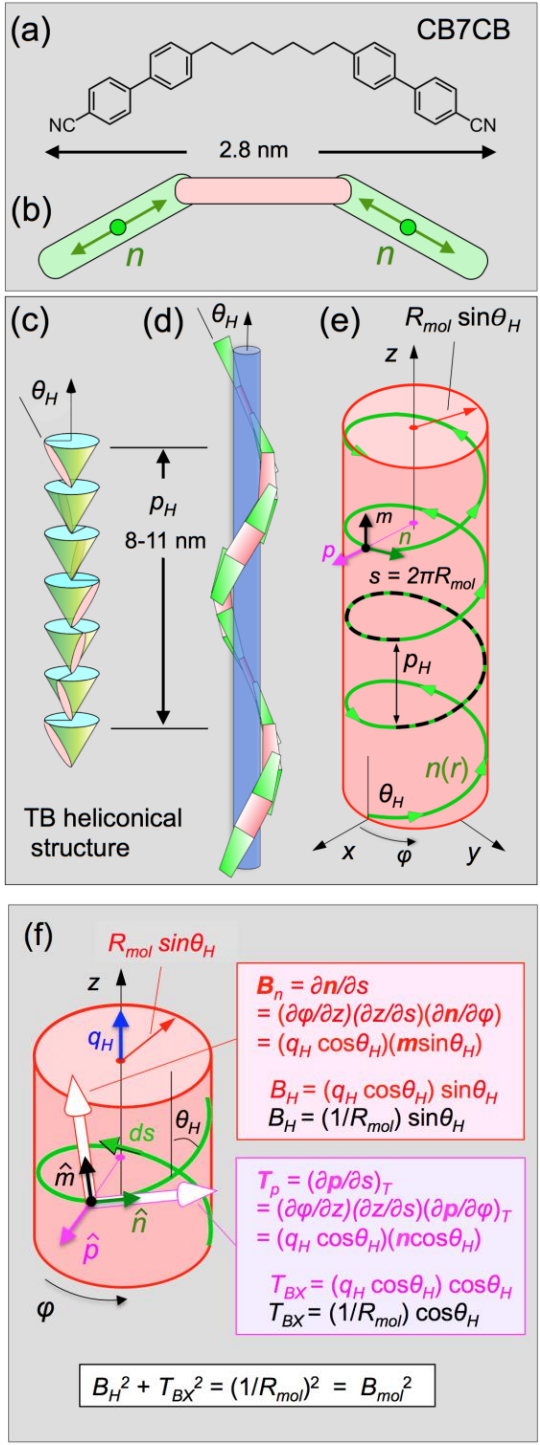

Figure 1: *a*) CB7CB, the bent molecular dimer studied. *b*) Bent rod representation of CB7CB, showing its two contributions to the director field, unit vector *n*(*r*). Schematic structures of the twist-bend nematic (TB) phase, showing *c*) the precession of the director orientation on the cone of angle $\theta_H$, taken for purposes here to be the tilt of a principal axis of the dielectric tensor. *d*) Helical winding of the director in the TB phase. At each level, the indicated orientations fill the x-y plane. *e*) Geometry of the helical path of the contour line that locally follows the orientation of *n*(*r*). The distance along the contour is $s(\varphi)$. A physical constraint, first reported in this paper, of the TB structure in CB7CB is that the cylinder radius varies with cone angle $\theta_H$ as $\sin\theta_H$, such that the length of contour $s(2\pi)$ for one pitch of the helix (dashed black line) is independent of $\theta_H$ and always given by $s(2\pi) = 2\pi R_{mol}$, where $R_{mol}$ is the bend radius of curvature of an extended CB7CB molecule. *f*) Geometry of reorientation on the director contour line. Of relevance to the TB phase is $B_n(r)$ the bend rotation of *n*(*r*) about auxiliary vector *m*(*r*), and $T_p(r)$, the twist rotation of the biaxial vector *p*(*r*) about *n*(*r*). The magnitudes of these deformations, $B_H$ and $T_H$ are uniform in space and, under the constraint noted above, satisfy the condition $B_H^2 + T_H^2 = 1/R_{mol}^2$, describing the exchange of director bend for biaxial twist as $\theta_H$ is decreased.

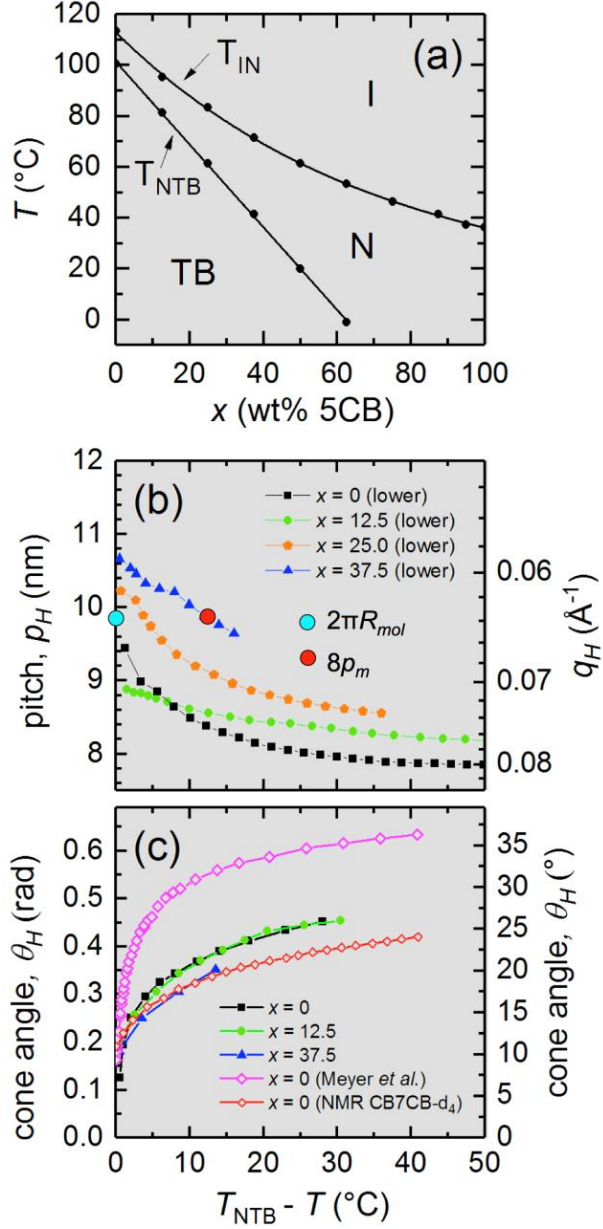

Figure 2: *a*) Phase diagram of the CB7CB/5CB mixtures vs. weight %, $x$, and temperature, $T$, exhibiting isotropic (I), nematic (N), and twist-bend (also termed heliconical) nematic (TB) phases. *b,c*) Helix pitch, $p_H$, obtained by resonant soft x-ray scattering (RSoXS), and optical cone angle, $\theta_H$, determined from birefringence measurements, of the heliconical structure in the TB phase vs. $x$ and $T_{NTB} - T$, where $T_{NTB}$ is the N–TB phase transition temperature. Birefringence [36] and NMR data [37] from literature sources are also included. In CB7CB, the helix pitch near the transition is found to be $p_H \approx 2\pi R_{mol}$ (blue dot in *b*)) where $R_{mol}$ is the bend radius of curvature of an extended CB7CB molecule. Diffusive non-resonant x-ray scattering indicates periodic segmentation along the helix of spacing $p_m \approx 1.25$ nm. The pitch near the transition has eight such segments (red dot in *b*)).

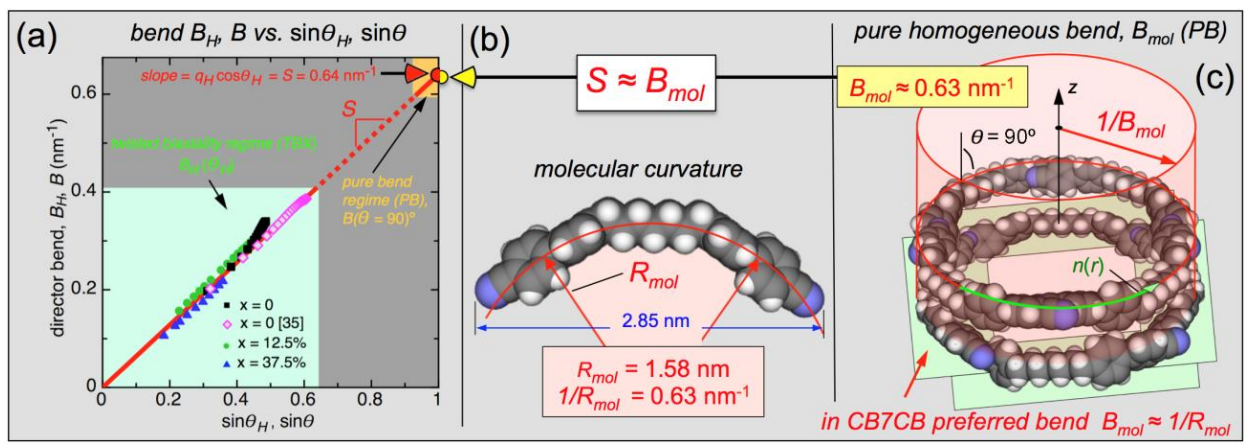

Figure 3: *a*) Director bend deformation magnitude $B_H(\theta_H) = (q_H\cos\theta_H)\sin\theta_H$ calculated from the $q_H$ and $\theta_H$ data of Fig. 2. The data lie closely to a straight line through the origin, indicating that $q_H\cos\theta_H$ = slope, $S$ = 0.64 nm$^{-1}$. Changing $x$ or $T$ just moves the points along the line. *b*) The $B_H(\theta_H)$ data of *a*) can be related to the molecular shape of CB7CB by noticing that $S$ is nearly equal to its inverse molecular radius of (bend) curvature $1/R_{mol}$. obtained by fitting atomic centers to a circle. This suggests that the extrapolation to $\theta$ = 90º describes the state of maximum intrinsic bend, obtained by putting the molecules into the state of pure homogeneous bend (PB) as in Fig. 3c, packing them while energetically pinning them to a cylinder of variable radius on which they can seek their intrinsic bend curvature $B_{mol}$. For CB7CB, which has a nearly circular shape, $B_{mol} \approx 1/R_{mol}$ and the pitch data of *a*) are described by $B_H = B_{mol}\sin\theta_H$. This relation elegantly connects macroscopic helix characteristics at small $\theta_H$, where the TB structure is mostly biaxial twist and has little bend, to the bent molecular shape.

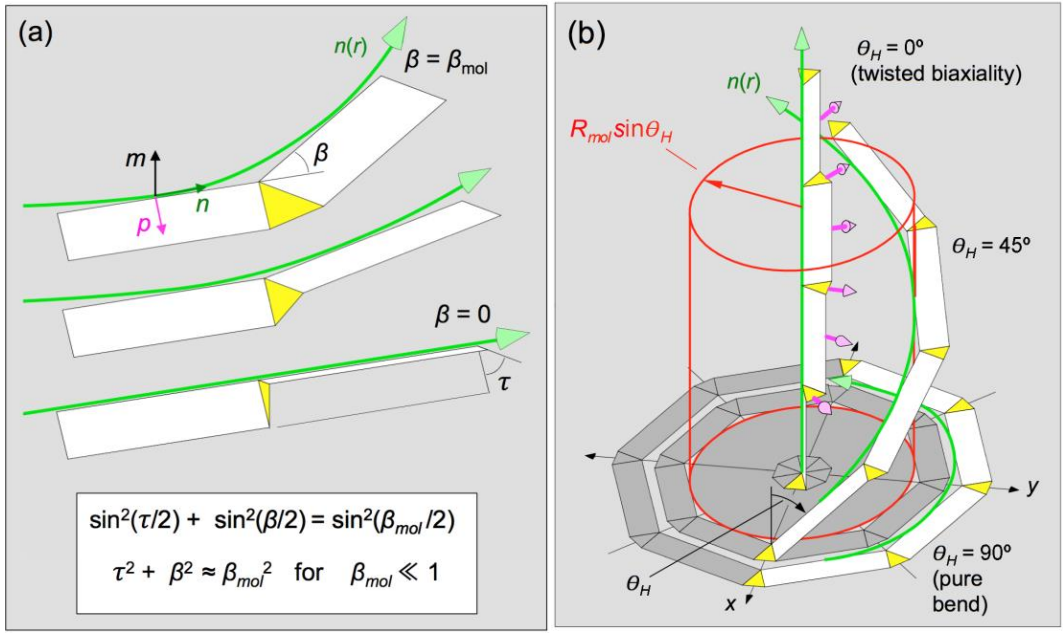

Figure 4: A polygon chain which models an elastic band that freely interconverts between director bend and biaxial twist. This chain quantitatively embodies the geometry of the TB heliconical state as manifest in the relationship $B(\theta_H) = (1/R_{mol})\sin\theta_H$ from Fig. 3a, and motivates our picture of the TB phase as an assembly of sterically stabilized oligomeric chains. *a*) In this geometry, the rigid triangular and rectangular plates form a chain by sharing common edges which are flexible hinges, enabling the chain to twist if its bend is reduced as sketched and described geometrically by the relationship indicated. *b*) The polygon chain is attached to a helical rod that is flexible but of fixed length $2\pi R_{mol}$, and which can be made to change its pitch by sliding its upper end along *z*. Here, only a half-period of the helix is drawn. The red cylinder changes radius as $\sin\theta_H$ to keep the rod length constant. In its flattened state ($\theta_H = 90°$) the chain models directly the PB regime in Fig. 3c, in this case with eight segments and eight 45° bends. If $\theta_H$ is reduced then this structure rigorously maintains the conditions $q_H(\theta_H)\cos\theta_H = 1/R_{mol}$, and $B(\theta_H) = (1/R_{mol})\sin\theta_H$, *i.e.*, it moves on "the line", as the data in Fig. 3a. In the fully stretched out state ($\theta_H = 0°$), the twisted biaxiality (TBX) regime, the director bend has disappeared to be replaced by twist of the biaxial vector *p* (magenta arrows), with a pitch $p_H = 2\pi R_{mol}$ mediated in eight twist steps of 45° each, as suggested by non-resonant x-ray scattering data.

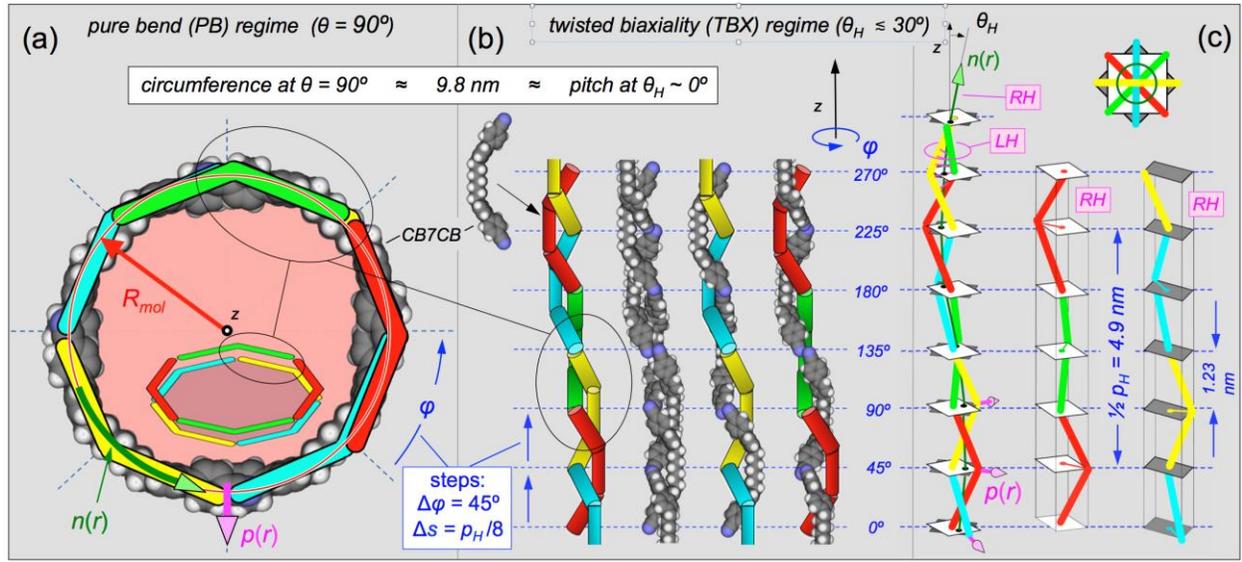

Figure 5: Realization of the polygon chain model for discrete molecules, modeling CB7CB by bent rods. Oligomeric chains are formed in both the PB and TBX regimes by the brickwork tiling of half molecules, as indicated by the blue dashed lines, the short-range periodicity in diffuse non-resonant x-ray scattering (Supplementary Figs. S11 and S13). Each segment contains a pair of half-molecules. This tiling is stabilized by the entropic association of molecular ends and CB7CB's flexible center. The constraint of the bend data and the polygon chain model is that the helix pitch $p_H$ at small $\theta_H$ in the TBX regime is the circumference of the circle in the PB regime. Since $p_H$ at small $\theta_H$ is 8 segments in length, the change in $\varphi$ per segment is taken to be 45º in both the PB and TBX regimes. *a*) PB regime showing the brickwork tiling of molecules as in Fig. 3c: two chains of 45º bent molecules (red-green, yellow-cyan) forming eight segments of half-molecule pairs, with angular bend jumps of $\Delta\varphi = 45º$ between each pair, and a $\Delta\varphi = 45º$ phase difference in the orientation of the two chains. *b,c*) Oligomeric chain structure showing bent-rod and molecular models of the brickwork tiling in the TBX regime: two chains of 45º bent molecules (red-green, yellow-cyan) forming eight segments of half-molecule pairs, with angular biaxial twist jumps of $\Delta\varphi = 45º$ between each pair, and a $\Delta\varphi = 45º$ phase difference in the orientation of the two chains. The two right-handed (RH) chains associate to form a left-handed (LH) double helix. *c*) For uniaxial half-molecules, the optical polarizability of a given segment can be obtained geometrically. The optical cone angle is $\theta_H \sim 10º$ for the structure drawn, the case where the molecular planes are untilted.

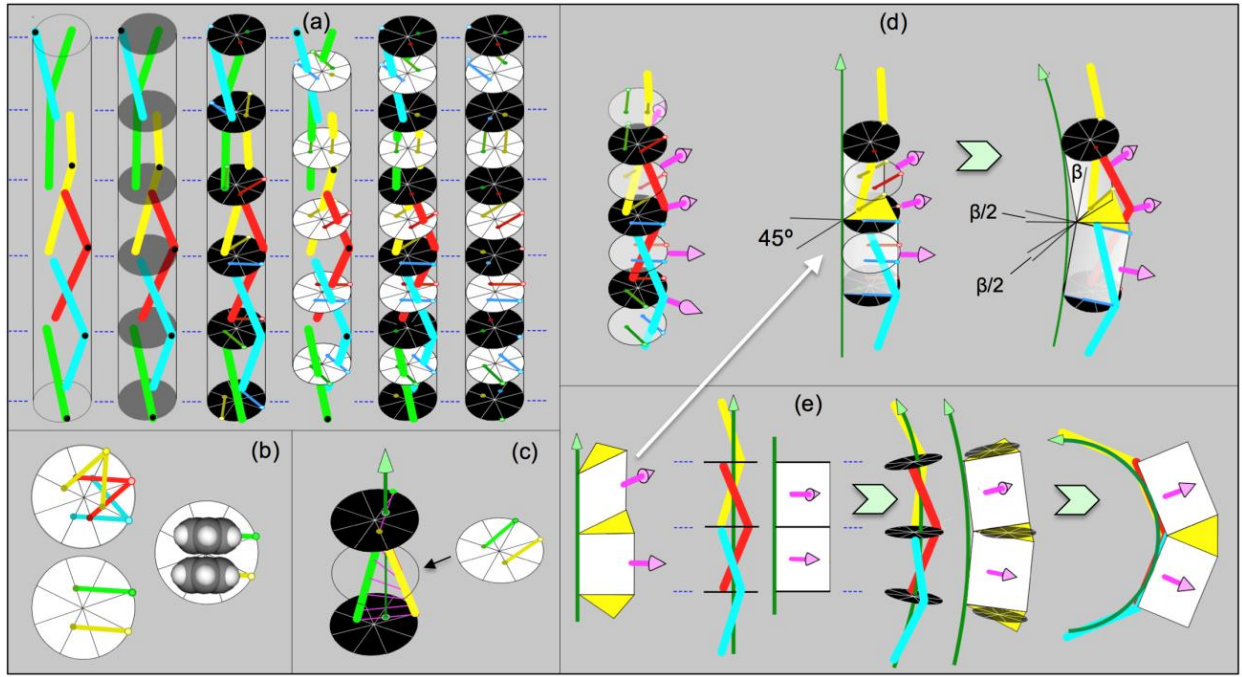

Figure 6: Representation of the TBX regime oligomeric chain structure for $\theta_H \sim 0$. *a*) White circles show the projections onto the x-y plane of half molecules in each segment, and the black circles the projections of half molecules on to the interfaces between segments. *b*) The planes of the 45º bent molecules are tilted from $z$ by an angle (~10º) such that the half-molecule projections on the x-y plane are separated by $\Delta\varphi = 45º$, matching the reorientation in successive segments (cyan, red, yellow). This results in parallel chain projections within the segments (yellow, green). This drawing is proportioned with respect to the diameter of the chains according to the molecular volume of 0.76 nm$^3$. The segments are 1.23 nm along $z$ and 0.88 nm in diameter, that of the black and white circles. The phenyl ring profiles are to the same scale *c*) When biaxial optical polarizabilities with parallel projection are added, they give an untilted biaxial average ($\theta_H \sim 0$). *d,e*) Mating of the polygon chain model with the brickwork oligomeric chain. The projections of the half-molecules onto an interface form a 45º angle that matches that of the yellow triangles in Fig. 4. If the oligomeric chain is bent, its black disc, representing the interface between two segments, tilts, but remains parallel to the yellow triangle. Thus, in the oligomeric chain, even as small $\theta_H$, the condition $q_H\cos\theta_H = 1/R_{mol}$ is enforced, keeping $B_H$ in Fig. 3a on "the line".

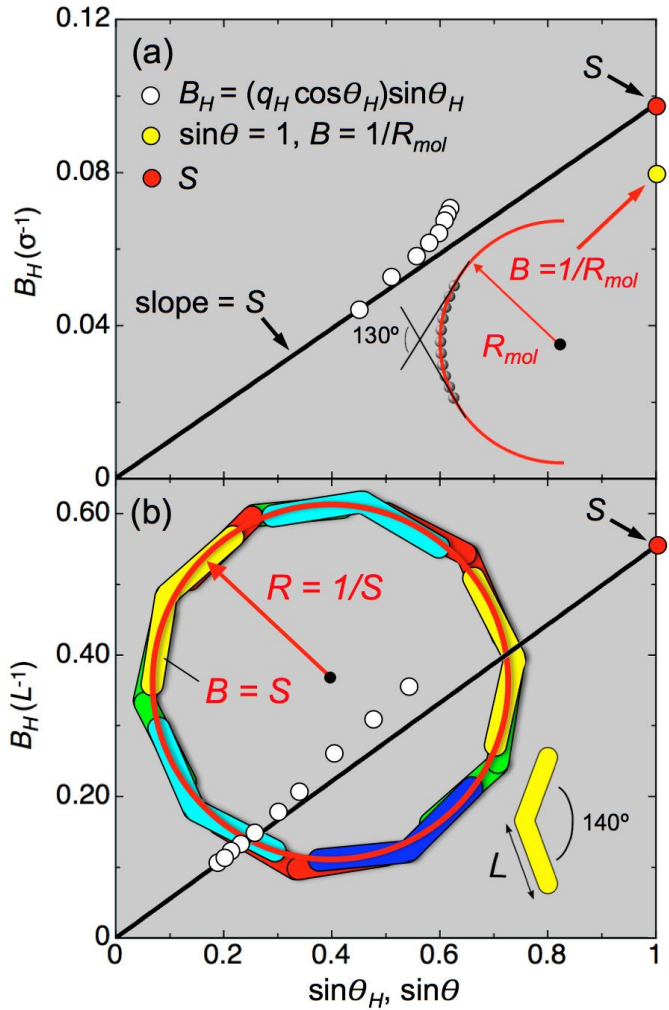

Figure 7: $B(\theta_H)$ vs. $\sin\theta_H$ data, analogous to those of Fig. 3a, for two model statistical mechanical systems of interacting bent particles. The black lines are drawn to match the bend data at the smallest $\theta_H$ *a*) Monte Carlo simulation of TB ordering of steric particles made by rigidly chaining 11 hard spheres in the form of circular arcs [51]. The length scale is the sphere diameter, $\sigma$. The molecular curvature construction of Fig. 3b is also sketched, giving a molecular bend $B_m = 1/R_{mol} = (2\pi \cdot 50)/(360 \cdot 11) = 0.079/\sigma$, plotted as the yellow dot at $\theta = 90°$. $B_m$ is comparable to the slope of the line, as was found for CB7CB in Fig. 3. *b*) Generalized Maier-Saupe mean-field theory of the TB ordering of bent rods in which the rod halves were considered as independent interaction centers [52]. Also sketched is the PB construction of Fig. 3c, where R is simply set to $R = 1/S$, i.e., the bend $B$ in the circular construction is set to $S$, the bend on the line extrapolated to $\theta = 90°$. It seems remarkable that a mean-field model can capture this geometry.

# SUPPLEMENTARY INFORMATION

*TABLE OF CONTENTS*

*MATERIALS AND METHODS*

CB7CB (4',4'-(heptane-1,7-diyl)bis(([1',1"-biphenyl]-4"-carbonitrile))) was synthesized by two different methods as described previously [1]. We purchased 5CB (4-cyano-4′-pentylbiphenyl) from Sigma-Aldrich and used it as received. We prepared mixtures of CB7CB and 5CB of concentration $x$ (where $x$ is the weight percent of 5CB in a 5CB/CB7CB mixture) by several iterations of mechanically mixing the two compounds at 120°C (at which temperature they are both isotropic liquids), followed by centrifugation for ~ 1 minute.

We filled the liquid crystal mixtures into Instec 3.2 μm unidirectionally rubbed commercial cells and home-made untreated cells at high temperature in the isotropic phase by capillary action. We carried out polarized transmission optical microscopy (PTOM) on these cells, using a Nikon Eclipse E400 polarizing microscope equipped with an Olympus Camedia C-5050 Zoom digital camera. Cells were slowly cooled from the isotropic phase (at ~2°C min$^{-1}$) to the desired temperature using an Instec STC200D temperature controller.

We measured the birefringence of CB7CB and its mixtures with 5CB in the PLM as described above with a Zeiss Ehringhaus Rotary Compensator in Instec 3.2 μm unidirectionally rubbed commercial cells using the C spectral line (656.3 nm). We cooled the samples from the isotropic phase at -1°C min$^{-1}$ and maintained them at constant temperature while recording the birefringence. We searched for the best aligned region in the cell and took 3 independent measurements of the birefringence at each temperature and averaged them to obtain the result.

We investigated the TB pitch of the mixtures by FFTEM and RSoXS. The analysis in the main paper is based solely on the RSoXS data, since the FFTEM measurements exhibited a wide variability and more uncertainty, which is discussed below in the Supplemental information.

We prepared the samples for FFTEM experiments by sandwiching the LC mixtures between 2 mm by 3 mm untreated glass slides spaced by a several-micron-thick gap and observed the cell on a hot stage under a PLM. We heated the cells to ~120°C then slowly cooled into the desired phase. Once the sample had equilibrated to the desired temperature, we ejected the cell from the hot stage into a well of liquid propane which rapidly quenches the cell to $T < 90$ K. We then transferred the cell into a Balzers BAF-060 freeze-fracture apparatus under high vacuum with $T \sim 140$ K, where we mechanically fracture the cell by pulling the glass plates of the cell apart to expose the surface topography of the bulk. We then shadowed this fracture face by oblique evaporation of 2 nm of platinum to provide contrast to the topography of the exposed surface. A final coating of 25 nm carbon normal to the surface enhances the mechanical rigidity of the coating, thereby completing the replica with which we viewed the interfacial topography and imaged it in a Philips CM10 transmission electron microscope.

We performed resonant x-ray scattering experiments at beamline 11.0.1.2 at the Advanced Light Source, Lawrence Berkeley National Laboratory. In tuning the beam about the carbon k-edge resonance $E_{res} = 283.5$ eV, we found that the scattering signal was observable only for energies near $E_{res}$.

## *PHASE BEHAVIOR*

We observed the N–TB phase transition on the various mixtures in untreated and planar cells up to the $x = 62.5$ mixture using DSC (Fig. S1) and PLM (Fig. S2). Below the N–TB transition, in mixtures of concentration $x = 0.0 – 50.0$, TB domains nucleate from the nematic phase and grow and coarsen until the phase transition is complete. Figs. S2a,b show N/TB phase coexistence in $x = 0.0$ and $x = 25.0$ samples, respectively. While director fluctuations are clearly visible in the nematic phase, they are not detectible in the TB phase. For the $x = 62.5$ mixture, the TB phase nucleates rapidly leaving the smooth, undulation free TB texture in the cell, as shown in Fig. S2c. When the 5CB concentration is further increased to $x = 75.0$, we observe the nematic director fluctuations freeze out at ~ –20°C, with no other observable indications of a phase transition to the TB phase down to ~ –100°C (Fig. S2d).

In the $x = 12.5$ and $x = 25.0$ mixtures, the TB phase exhibits the stripe textures typical of neat CB7CB (Fig. S2), these stripes being due to an undulation instability [2] caused by the shrinking of the TB pitch on cooling [1]. On reducing the temperature of the mixtures with lower 5CB concentrations to only several degrees below the N–TB transition, we observe the stripe texture form. These stripes appear and then disappear as the temperature settles, until they relax away nearly completely after several minutes, and a smooth texture of uniform birefringence remains, shown in Figs. S3a,b for the case of $x = 37.5$. On further cooling into the TB phase, we observe the stripes fill the cell once again (Fig. S3c), but only a fraction of them relax back into the uniform state this time (Fig. S3d). In $x = 37.5$, the stripes begin to persist in the phase at ~5°C below the N–TB transition. This behavior is likely due to a dramatic increase in the viscosity of the TB phase on decreasing temperature [3], where at a certain temperature the undulations cannot anneal away on the timescale of minutes. On increasing the 5CB concentration, however, the TB phase is more fluid. We find that in the $x = 62.5$ mixture, ~12°C below the N–TB transition, we still observe few stripes which then relax into the uniform state in a matter of several seconds, indicating that the addition of 5CB fluidizes the phase sufficiently so that it can relax into an undulation-free director field configuration [4] (Fig. S2c). This demonstrates the utility of making mixtures of 5CB with CB7CB to obtain a uniform and undulation-free, well-aligned TB phase.

We plot the latent heat released at the I–N and N–TB transitions as a function of 5CB concentration in Fig. S4. The latent heat of the I–N transition increases with increasing 5CB concentration in the mixtures. The linearity of the latent heat released indicates that CB7CB and 5CB are nearly ideally miscible in the N phase. We note that the latent heat of the I–N transition in CB7CB is about half that of 5CB, likely because the two "arms" of CB7CB are linked through an alkyl chain which reduces the conformational degrees of freedom available to the molecule, and amounts to a halving of the latent heat of transition of CB7CB with respect to that of 5CB at the I–N transition.

The latent heat released in the N–TB transition, on the other hand, decreases with increasing 5CB concentration (Fig. S4). This decrease is faster than expected from freezing point depression, based on the relative proportions of CB7CB and 5CB present in the mixtures. When the concentration reaches $x = 37.5$, the latent heat of the N–TB transition is barely detectable in the DSC plots (Fig. S1). This behavior is consistent with a calorimetric study on 5CB/CB9CB mixtures [5] in which a steady decrease in the latent heat of the N–TB transition with increasing 5CB concentration was observed until concentrations above 40% 5CB. The decrease in specific heat associated with the N–TB transition of the 5CB/CB7CB mixtures along with the PLM observations indicate that the first-order nature of the transition weakens with increasing 5CB concentration.

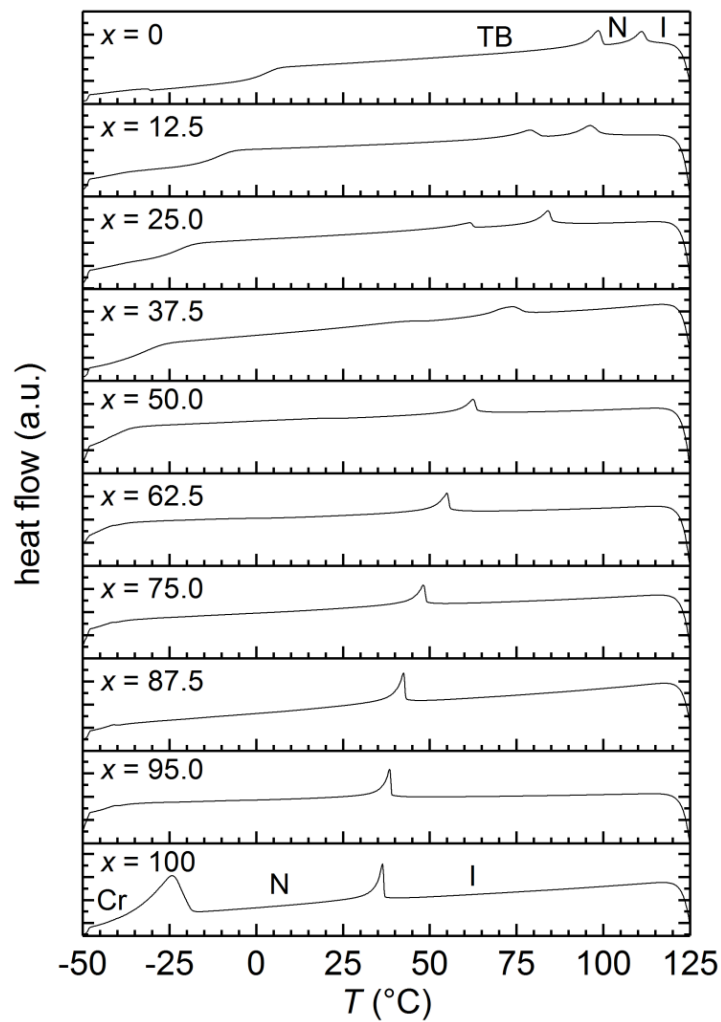

FIG. S1. DSC cooling curves of neat CB7CB, neat 5CB, and their mixtures obtained at a cooling rate of –10°C/min. In neat CB7CB and in mixtures with up to $x = 37.5$, we see peaks corresponding to both the I–N and N–TB transitions. Mixtures with higher 5CB concentrations exhibit only an I–N transition peak.

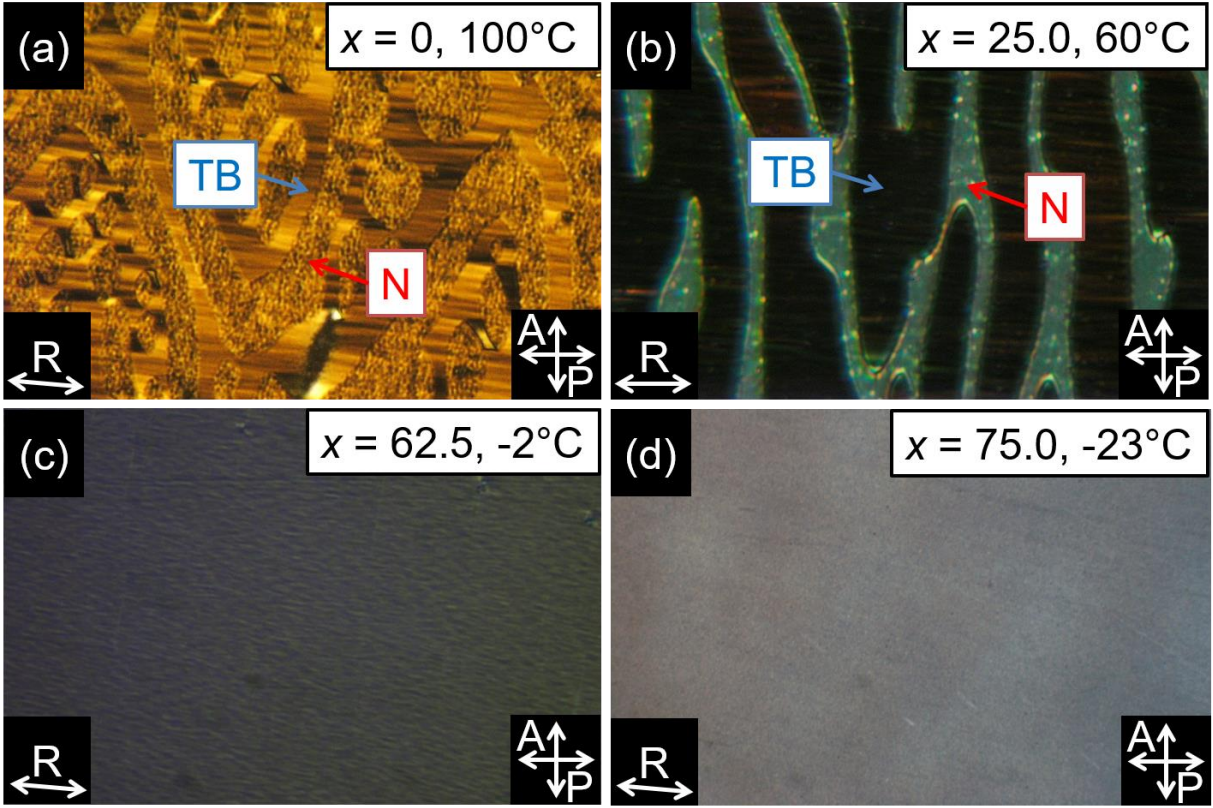

FIG. S2. Optical textures of neat CB7CB and 5CB/CB7CB mixtures near the N–TB transition in unidirectionally aligned planar cells. In *a)* $x = 0$ and *b)* $x = 25.0$, we see N–TB phase coexistence when the samples are cooled slowly from the nematic, confirming the first order nature of the N–TB phase transition in these samples. As we increase the 5CB concentration, the temperature ranges over which we observe phase coexistence becomes narrower. In the *c)* $x = 50.0$ and *d)* $x = 62.5$ mixtures, a very weak stripe texture appears when cooled several degrees below the N–TB transition, with no observable phase coexistence.

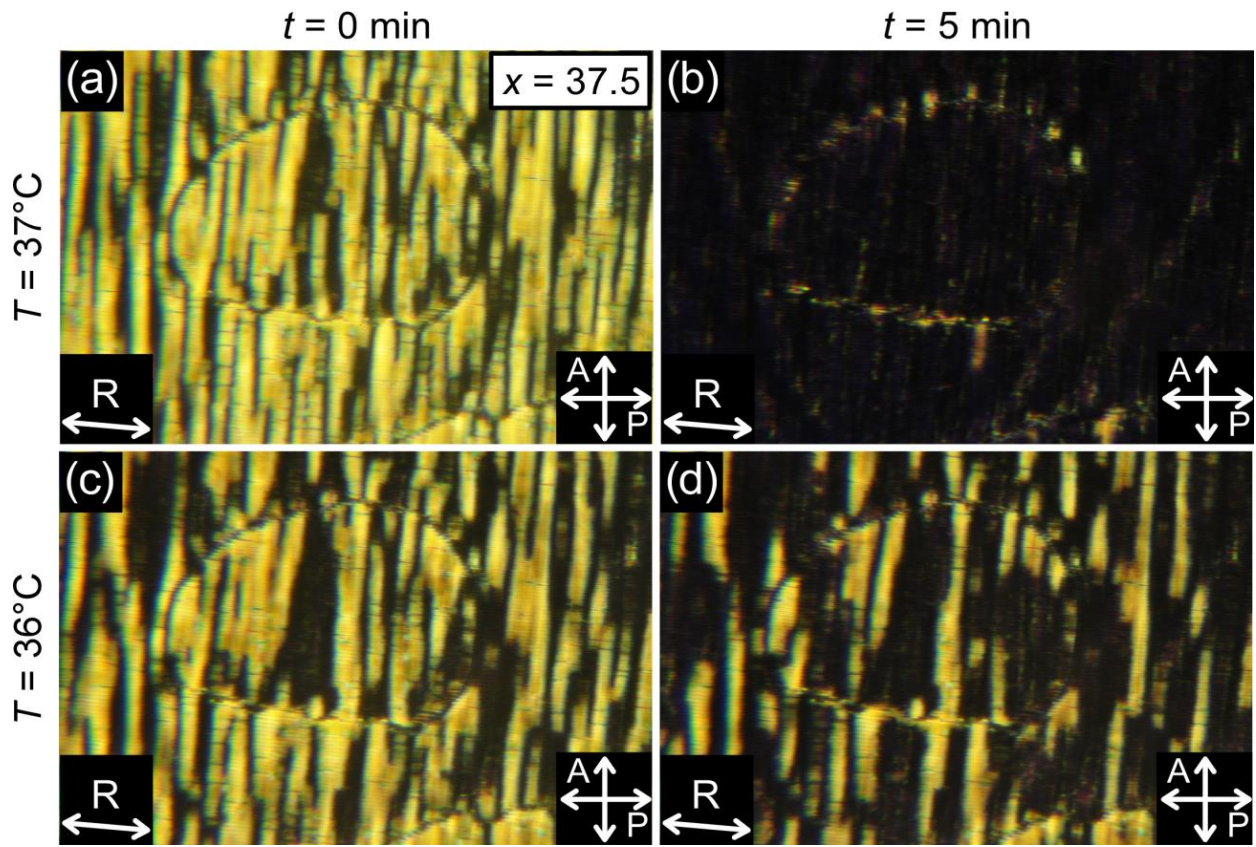

FIG. S3. Annealing of stripe textures in the $x = 37.5$ mixture. On cooling from 38°C to 37°C, a transient stripe texture appears *a)*, which anneals into a (mostly) uniformly aligned state after 5 minutes *b)*. On further cooling from 37°C to 36°C, stripes appear once again *c)*. Because of the increased viscosity of the mixture on cooling, the texture does not completely anneal into the uniform ground state *d)*. Further cooling produces more stripes which persist, until the cell is filled with them.

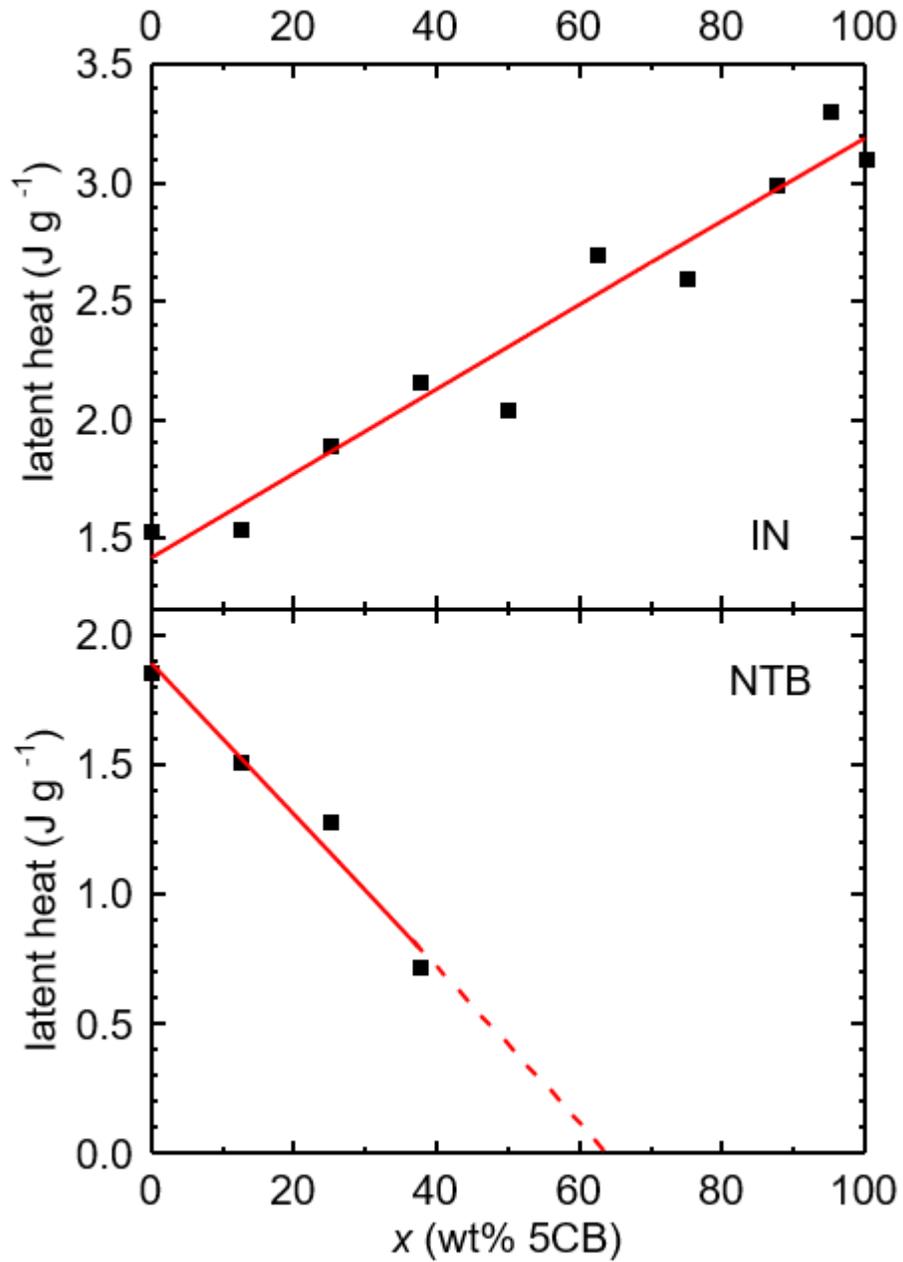

FIG. S4. Latent heat of the I–N and N–TB transitions in mixtures of CB7CB and 5CB. The latent heat of the I–N transition increases continuously with 5CB concentration, reflecting the larger latent heat content of the I–N transition of neat 5CB. The latent heat of the N–TB transition in the mixtures decreases more quickly than expected based on the concentration of the components alone.

*DETERMINATION OF PITCH BY RESONANT SOFT X-RAY SCATTERING (RSOXS)*

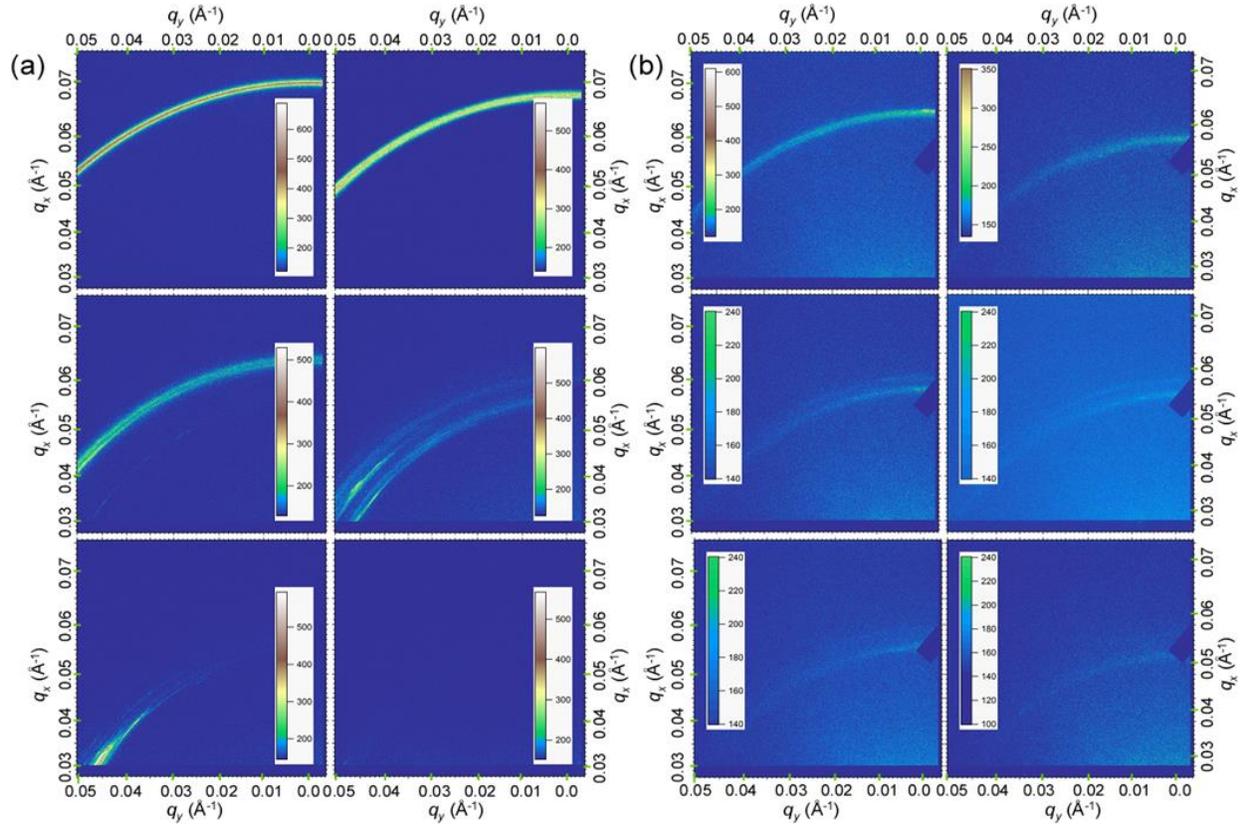

FIG. S5. RSoXS images of *a)* $x = 25.0$ and *b)* $x = 37.5$ mixtures taken on increasing temperature. The incident x-ray energy is near the carbon $K$ edge ($E_{res} = 283.5$ eV). For both mixtures, the scattering arc is relatively narrow and well-defined at low temperature. The diffraction arcs shift to smaller $q_H$ and broaden on increasing temperature. The scattering arcs in $x = 37.5$ appear broader and more diffuse than those from the $x = 25.0$ mixture.

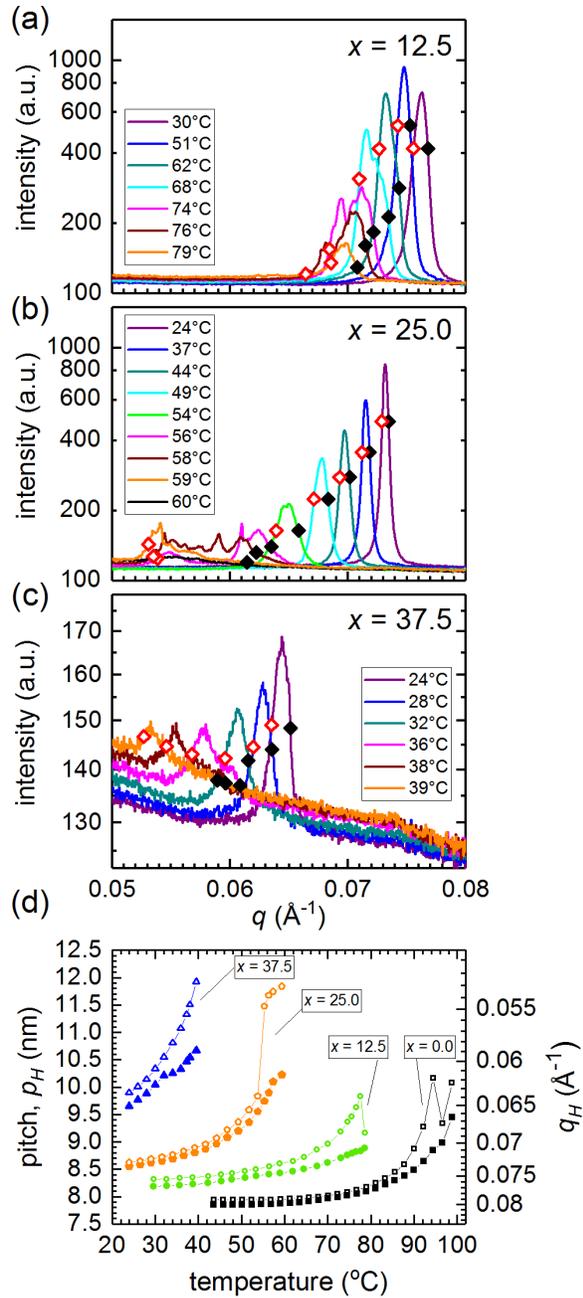

FIG. S6. Azimuthally averaged intensities $I(q)$ obtained from 2D RSoXS SAXS images such as those in Fig. S5. The scans at different temperatures show the variation of scattering peak position and shape with temperature on heating for the *a)* $x = 12.5$, *b)* $x = 25.0$, and *c)* $x = 37.5$ mixtures. The open red and solid black diamonds denote the upper and lower limits of the intensity distribution, respectively, showing its broadening in $q$ with increasing $T$. The pitches corresponding to the limits are calculated from $p_H = 2\pi/q_H$ and plotted in *d)*. The lower limit in $q$ depends on the maximum dilative stress in the sample texture. The upper limit in $q$ remains the same for different temperature scans, so it is taken to give the measurement of $q_H$, corresponding to the lower limit of $p_H$ (solid points in *d)*). These are plotted in Fig. 2b and used to calculate the heliconical bend magnitude, $B$.

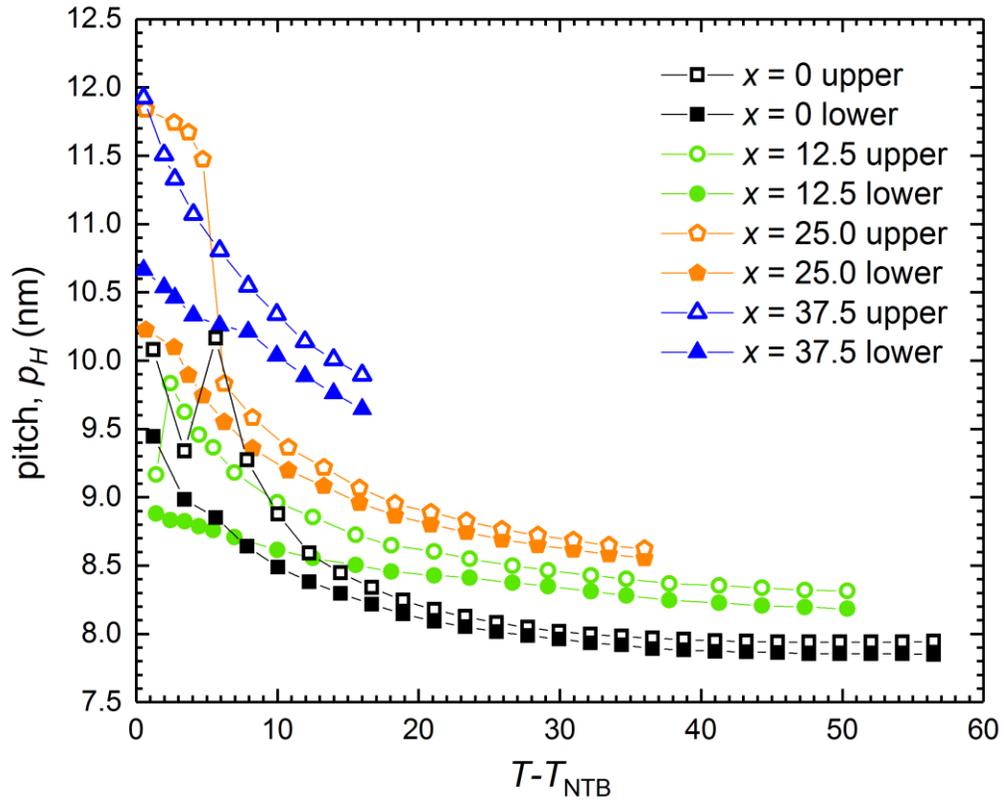

FIG. S7. Pitches $p_H$ corresponding to the upper and lower limits of the RSoXS scattered intensity distribution $I(q)$, plotted as a function of temperature relative to the N–TB transition temperature for CB7CB and its mixtures with 5CB.

## *DETERMINATION OF OPTICAL CONE ANGLE BY BIREFRINGENCE MEASUREMENT*

We measured the birefringence as a function of temperature for neat CB7CB and its mixtures with 5CB in the N and TB phases to determine the TB cone angle $\theta_H$. In the uniaxial N phase the birefringence of the neat CB7CB and the mixtures (Fig. S8) is fitted with the Haller formula [6]

$$\Delta n = \Delta n_0 (1 - T/T_{\text{NI}})^\beta, \tag{1}$$

where $T_{IN}$ is the I–N transition temperature, $\Delta n_0$ is the birefringence extrapolated to $T_{IN} = 0$, and $\beta$ is a parameter controlling the variation in the order parameter with temperature.

We can obtain the effective molecular bend angle ($\pi$–$2\chi$) of the CB7CB molecules in the uniaxial nematic phase with $\Delta n_0(CB7CB)$ from eqn. (1) and a formula that was first described by Meyer *et al.* [7], in which we assume that the difference in $\Delta n_0$ between CB7CB and 5CB is due to the bent shape of CB7CB:

$$\frac{1}{2}(3\cos^2\chi - 1) = \frac{\Delta n_0(\text{CB7CB})}{\Delta n_0(\text{5CB})}. \tag{2}$$

The results are displayed in Table S1. The bend angle of neat CB7CB in the uniaxial nematic phase is found to be 118.5°, which corresponds quite well to a statistical-mechanical calculation which shows that the distribution of bend angles peaks at 119–120° [8]. As we increase the 5CB concentration, the effective bend angle decreases only slightly.

Assuming the magnitude of the order parameter ($S(T)$) of CB7CB and its mixtures in the TB phase follows the same trend in temperature as that extrapolated from fitting the nematic phase, we can calculate the temperature-dependent heliconical tilt angle $\theta_H$ of the TB phase in neat CB7CB and its mixtures with 5CB by applying the following formula [9]:

$$\Delta n(T) = \Delta n_0 S(T)(3\cos^2\theta_H - 1)/2. \tag{3}$$

where the uniaxial nematic phase has a heliconical tilt angle of 0°. As shown in Fig S9, for both the neat CB7CB and its mixtures, after passing through the N–TB phase transition, the heliconical cone angle increases dramatically with decreasing temperature.

We compare the tilt angle as obtained from the analysis of the experimental birefringence data described above with data derived from birefringence measurement reported by Meyer *et al.* [7] and NMR data reported by Jokisaari *et al.* [10] (Fig. 2 of the main text). The discrepancy between our cone angle measurements and those of Meyer *et al.* appear to come down to the choice of region in which to take the birefringence measurement. Meyer *et al.* measure a single monochiral domain, which on cooling forms TB stripe textures that decrease the birefringence. Therefore, their birefringence measurement may overestimate the cone angle as the temperature decreases beyond the vicinity of the N–TB transition. We averaged over a larger well-aligned region in the cell, which generally will have a higher birefringence than a single given domain.

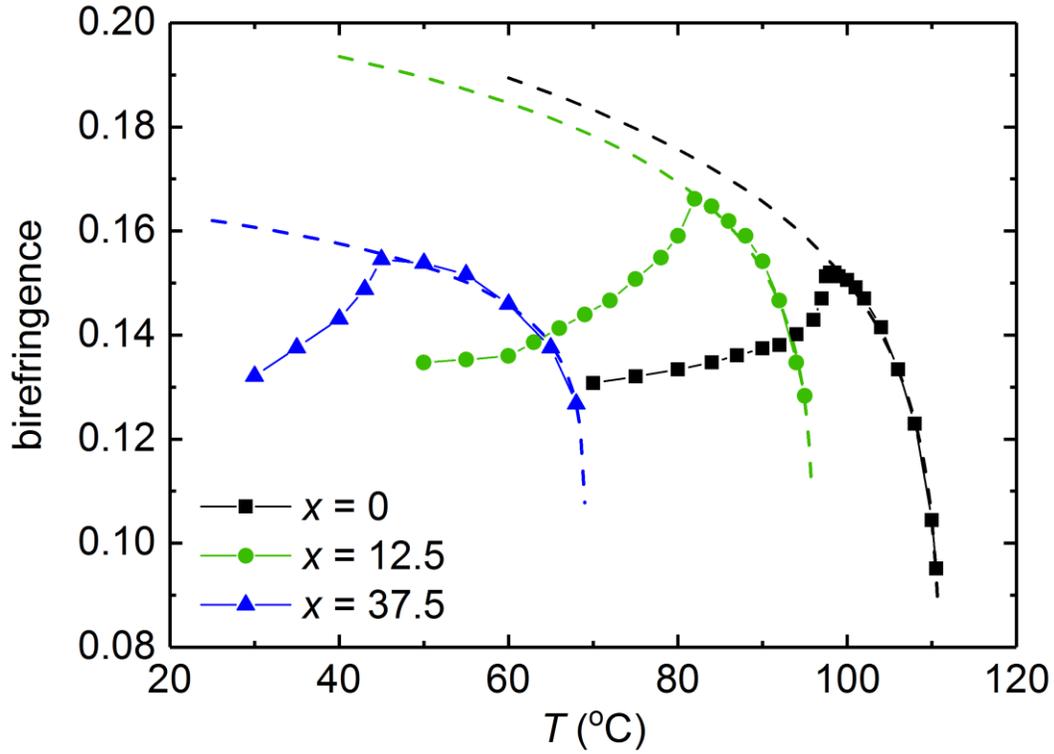

FIG. S8. Birefringence measurements of neat CB7CB and several of its mixtures with 5CB as obtained in POTM using a Berek optical compensator. The dashed curves are the fits to the Haller formula (Eqn. (1) above), using the data from the nematic phase.

TABLE SI. Fitting parameters from the birefringence data of neat CB7CB and its mixtures with 5CB.

| sample | $\Delta n_0$ | $\beta$ | $T_{NI}$ | bend angle $\chi$ (°) |
|---|---|---|---|---|
| x = 0 | 0.213 | 0.151 | 111.0 | 118.5 |
| x = 12.5 | 0.205 | 0.107 | 96.1 | 116.5 |
| x = 37.5 | 0.167 | 0.067 | 69.1 | 107.6 |
| 5CB (Ref. [11]) | 0.351 | 0.189 | 33.4 | – |

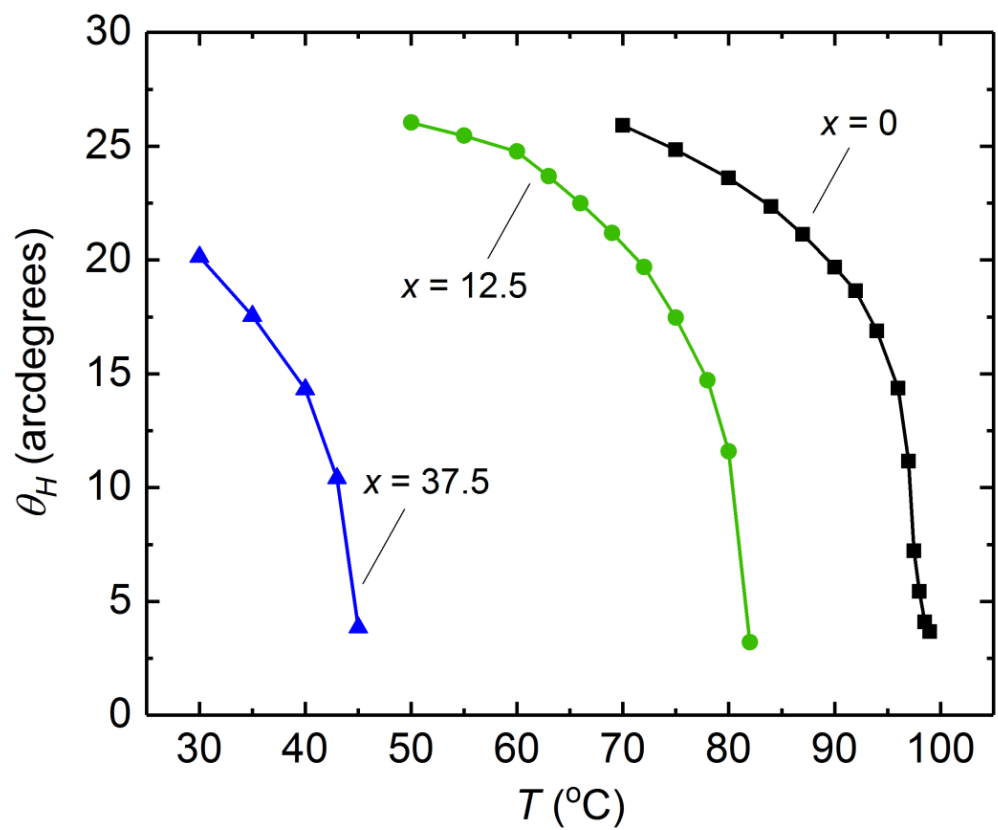

FIG. S9. Dependence of the heliconical cone angle $\theta_H$ on temperature for neat CB7CB and its mixtures with 5CB as obtained by birefringence measurements from a POTM using a Berek optical compensator.

## *BEND DEFORMATION IN THE TB PHASE*

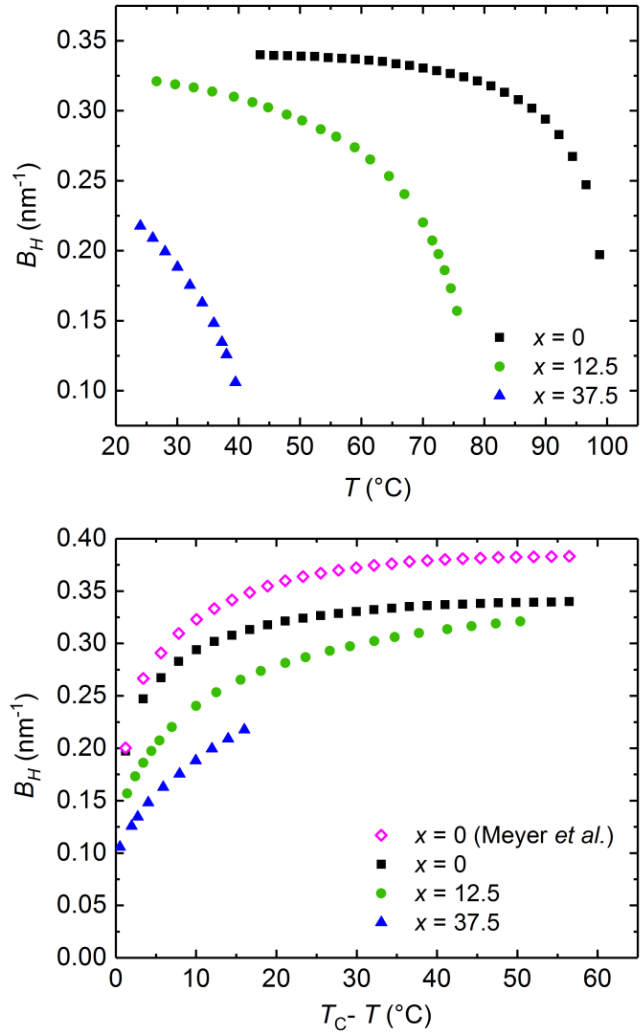

FIG. S10. The bend elasticity $B_H = (q_H \cos\theta_H)\sin\theta_H$ versus temperature ($T$) and the temperature shifted from the N–TB transition temperature ($T_{NTB} - T$).

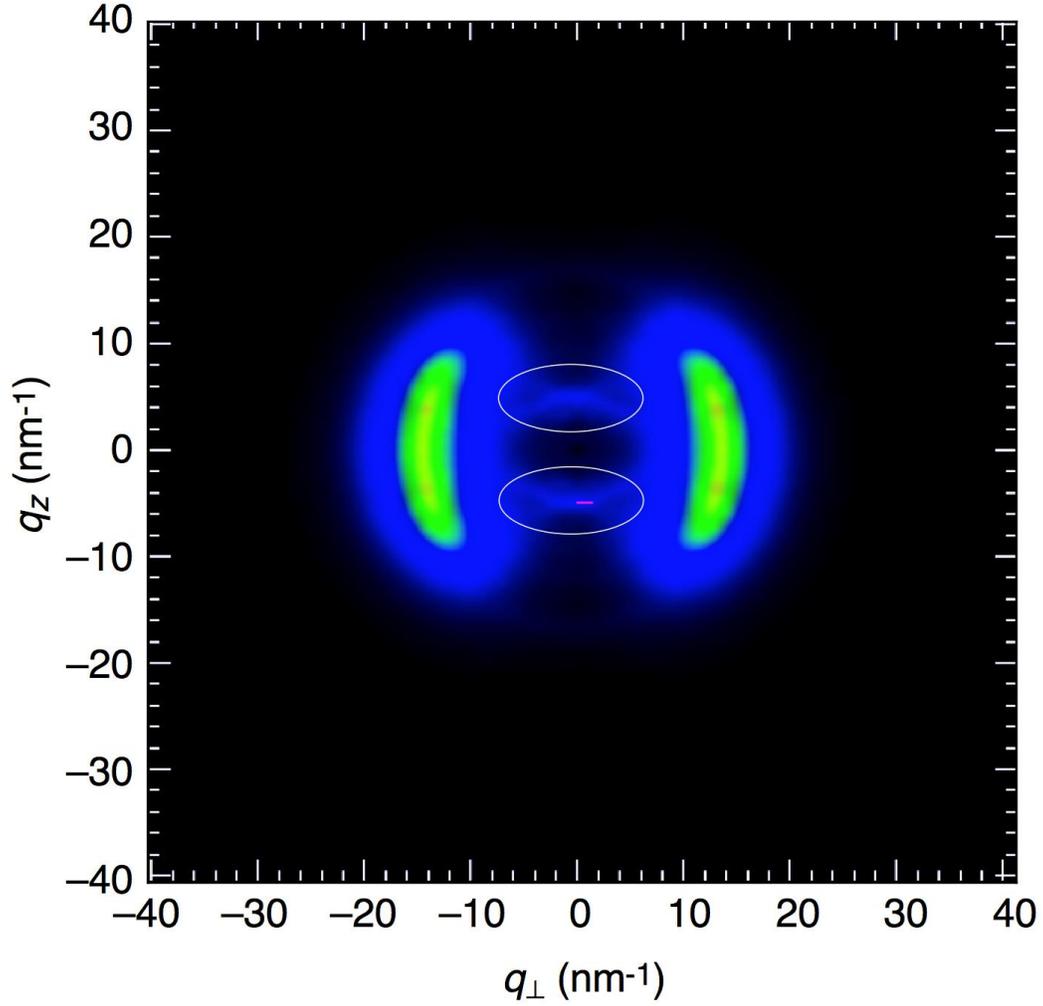

FIG. S11. Structure factor of the TB phase of CB7CB calculated from the molecular dynamic simulation data of ref. [12]. The white elliptical lines indicate the chain segment scattering, $I(q_z,q_\perp)$, with peaks at $q = \pm q_m = \pm 4.75$ nm$^{-1}$ ($d_m = 1.32$ nm). The half width in the $q_\perp$ direction indicates a correlation length $\xi_\perp \sim$ 0.5 nm in the direction normal to the axis, comparable to the diameter of a duplex chain. Correlation in the $q_z$ direction in $I(q_z,q_\perp)$ x-ray data is discussed in FIG. S12, and the value of $q_m$ from x-ray data in FIG S13.

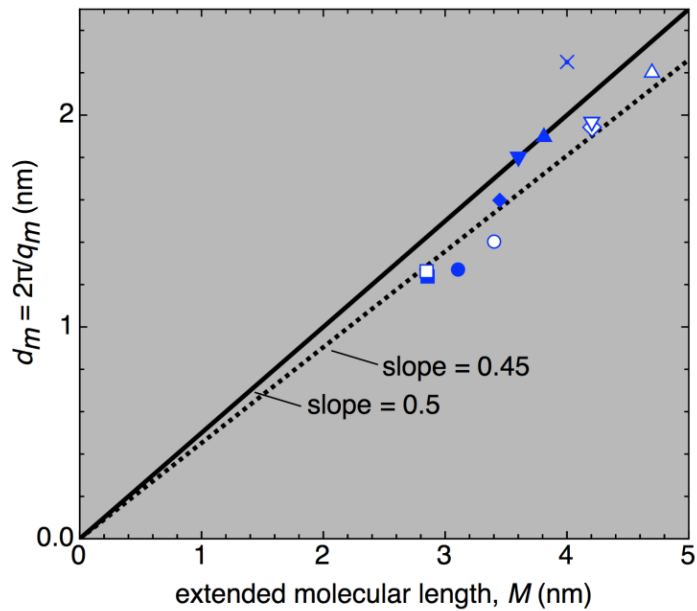

FIG. S12. Compilation of published measurements of spacings, $d_m$, calculated from the wavevector position of the on-axis diffuse non-resonant x-ray scattering peak of bent molecular dimers with flexible central links, plotted vs. their extended molecular length. The solid line has a slope 0.5 and the dotted line is a fit to the data. The plot shows a that $d_m$ is comparable to half the molecular length. This result supports the proposed duplex chain tiling, with the slope of 0.45 implying a 10% overlap of adjacent tail ends in the chain.

| symbol | reference number | full reference |
|---|---|---|
| ◇ | [13] | M. G. Tamba, et al., RSC Adv. **5**, 11207 (2015). |
| ◆ | [14] | K. Adlem, et al., Phys. Rev. E **88**, (2013). |
| △ | [15] | V. P. Panov, et al., Phys. Rev. Lett. **105**, 167801 (2010). |
| ✕ | [16] | R. J. Mandle and J. W. Goodby, Soft Matter **12**, 1436 (2016). |
| ● | [17] | V. P. Panov, J. K. Vij, and G. H. Mehl, Liq. Cryst. 1 (2016). |
| ☐ | [18] | E. Gorecka, et al., Liq. Cryst. **42**, 1 (2015). |
| ■ | [8] | M. Cestari et al., Phys. Rev. E **84**, 031704 (2011). |
| ▼ | [19] | A. Zep et al., J Mater Chem C **1**, 46 (2013). |
| ▲ | [20] | C. T. Archbold et al., Soft Matter **11**, 7547 (2015). |
| ▽ | [21] | W. D. Stevenson et al., ArXiv161201180 Cond-Mat (2016). |

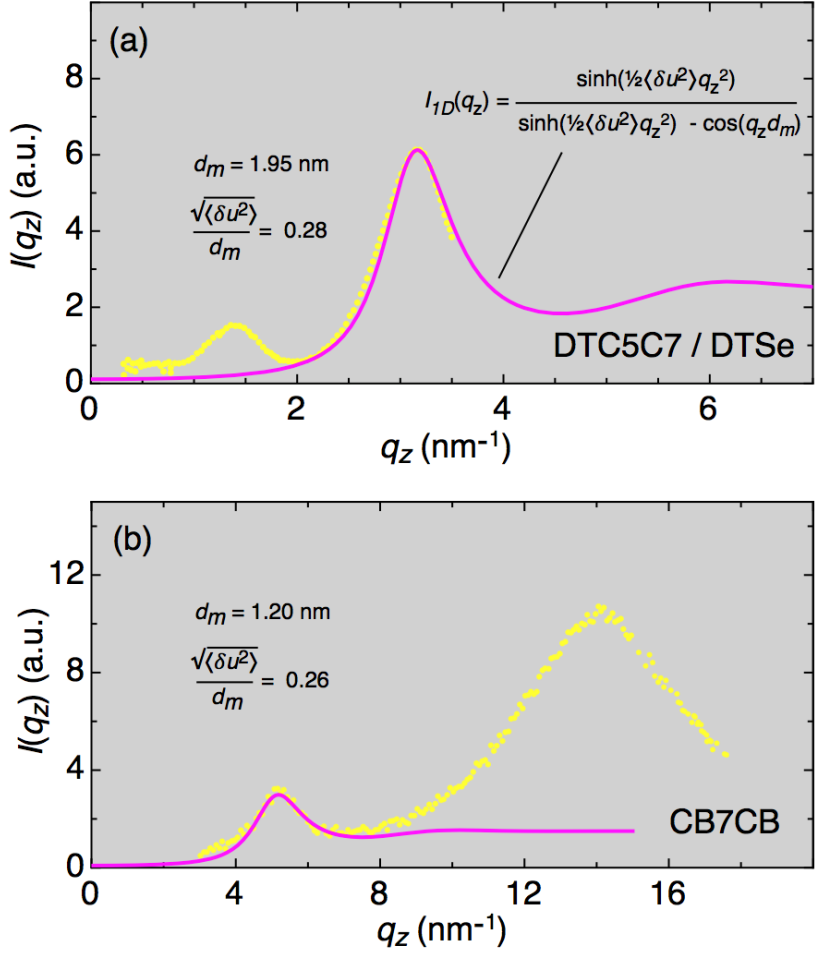

FIG. S13. Here we study quantitatively on-axis diffuse non-resonant x-ray scattering peak of bent molecular dimers with flexible central links in the TB phase in two materials where scans are published. This study is motivated by FIG. S12, which shows that the d-spacing observed is that expected for the half-molecule segment structure of the duplex tiled chins. We model the periodic segment structure as a one-dimensional chain of identical particles (segments) connected by harmonic springs. The corresponding structure factor, $I_{1D}(q_z)$, shown in *a)* [22,23], is fitted to the published scans from DTC5C7 [21] and CB7CB [8]. $I_{1D}(q_z)$ depends on three parameters: a multiplicative amplitude; $\langle \delta u^2 \rangle$, the mean-square relative displacement of nearest neighbor particles; $d_m$, the average interparticle spacing. The fits show that this simple chain 1D model can describe the segment periodicity peak well, with relative RMS displacement of the neighboring segments ~25% of their length, a reasonable value for such an assembly. DTC5C7/DTSe shows a subharmonic peak indicating a weak tendency for full molecular length ordering, that sharpens into a smectic layer reflection at low $T$. The broad peak in the CB7CB scan is that from the side-by-side molecular packing (e.g, highest intensity in FIG. S11), that shows up in this powder average.

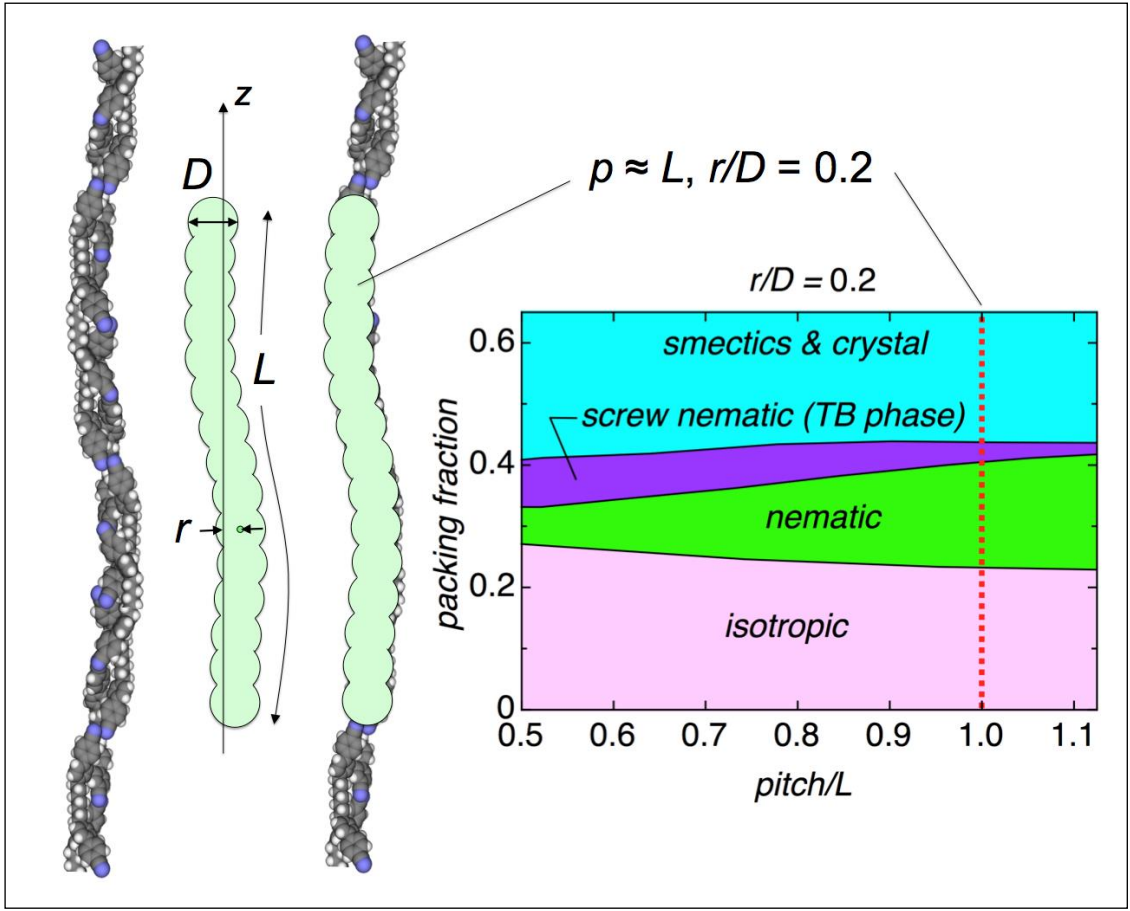

FIG. S14. Exploration of the 3D phase ordering of duplex tiled chains as a mechanism for the formation of the 3D heliconical structure of the TB phase, using the simulations of Kolli *et al*. [24] for the ordering of sterically interacting helical particles. The Kolli *et al.* particles are rigid assemblies of 15 hard spheres, each truncated to make the chain $10D$ long, in the form of a helix. Comparison of the steric shape of a duplex tiled chain of CB7CB with that of these particles shows that particles with radius of the helical deformation, *r,* such that $r/D = 0.2$ match the duplex shape well, with a length of about a single pitch.

The simulations equilibrated these particles as a function of packing fraction, generating the phase diagram shown for the case $r/D = 0.2$, where *r* is the radius of the helical deformation. The particles with $pitch/L = 1$ exhibit I, N, TB, and smectic phases with increasing packing fraction. The TB range decreases as *pitch/L* increases, narrowed at higher density by smectic ordering with layer thickness comparable to the particle length. Such smectic ordering would be suppressed relative to the columnar ordering of the TB phase by polydispersity in length of the chains, as would be expected in a self- assembly. The simulations show the feasibility of obtaining the TB phase is the duplex chains are sufficiently robust to interact sterically.

.

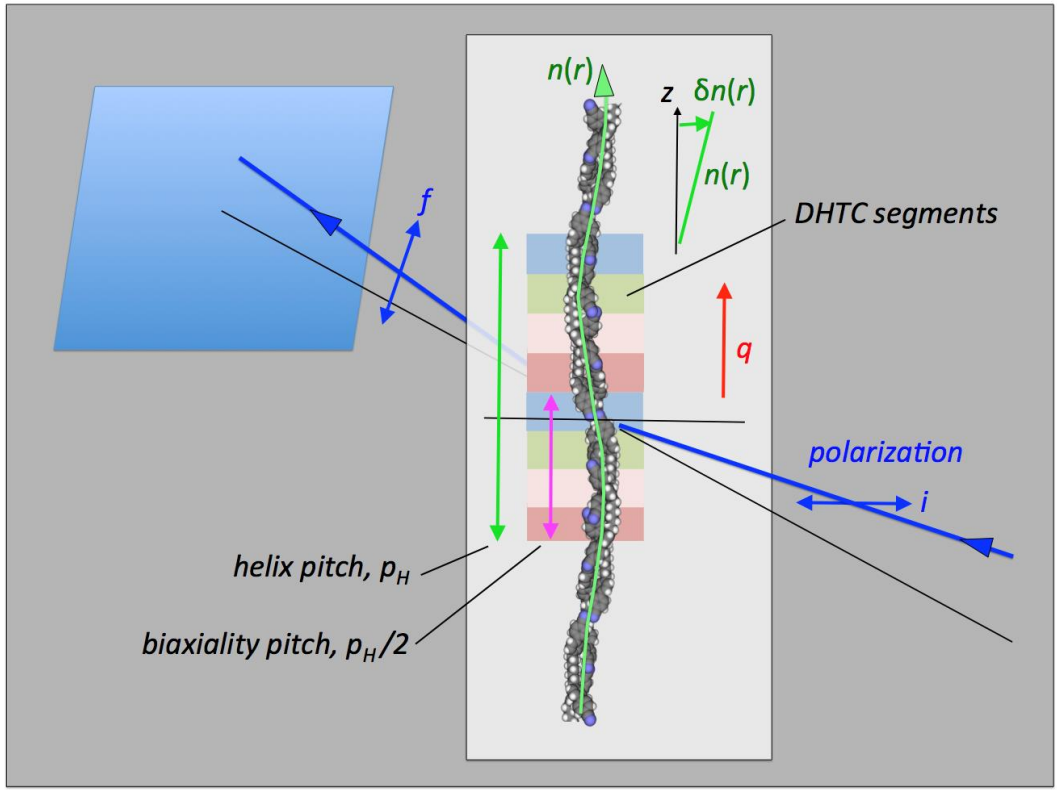

FIG. S15. RSoXS experimental geometry for scattering from a single DHT chain. Scattering from director rotation is depolarized and the field amplitude is linearly proportional to $\delta n(z)$, scattering at a wavevector $q = q_H = 2\pi/p_H$. Scattering from the periodicity of the biaxiality (pastel blocks) is polarized and at a wavevector $q = 2q_H = 4\pi/p_H$.

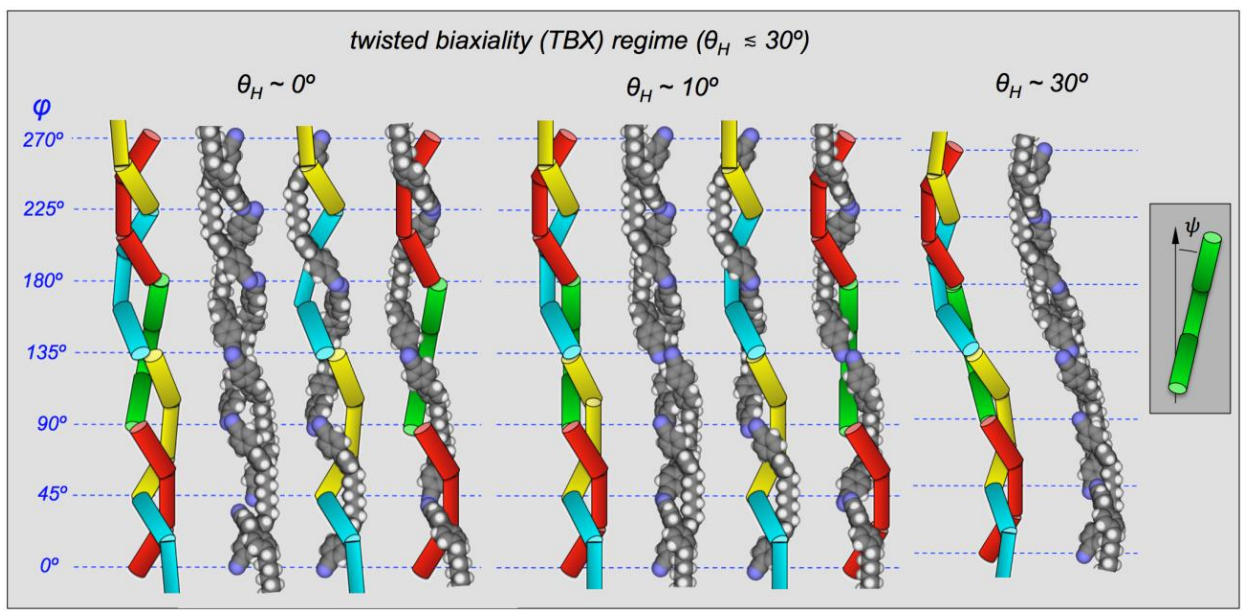

FIG. S16. Comparison of duplex tiled chains with different cone angles, $\theta_H$. In the $\theta_H = 10°$ case, shown in Figs. 5b,c, the tilt of the molecular planes is $\psi = 0$, as is clear from the drawings. This untilted case could occur at some particular temperature, in analogy to the unwinding of the helix in a chiral nematic at a particular temperature. As $\psi$ increases, $\theta_H$ is lowered, going to zero at about $\psi = 10°$, the case in Fig. 6. Larger $\theta_H$ values are obtained by helical deformation of the chain, perhaps the collective response of many chains once the 3D heliconical organization of the chains takes place, as in FIG. S14.

## *OBSERVATION OF THE HELICONICAL STRUCTURE AND PITCH MEASUREMENT BY FFTEM*

We have investigated the pitch $p_H$ of the TB phase of the mixtures by both freeze-fracture transmission electron microscopy (FFTEM) and RSoXS. The heliconical structure was first directly confirmed by FFTEM, but the FFTEM method exhibits a large inherent variability in the measured pitch values it gives, so that RSoXS is a much more reliable method for determining $p_H$. Thus, the analysis in the main paper is based solely on the RSoXS $p_H$ data. Here we discuss our data analysis process and the variability arising in measurements of $p_H$ using FFTEM.

Several dozen FFTEM images were made of the TB helix in each 5CB/CB7CB mixture to visually probe their nanoscale structure. FFTEM images of the mixtures exhibit nanoscale periodic modulations nearly identical to those found in the TB phase of neat CB7CB [12,25] (Fig. S16). In both neat CB7CB and in 5CB/CB7CB mixtures, neither stripe alternation nor half-order reflections in the Fourier transforms of the modulation patterns are present in the FFTEM images, which suggests that the observed periodicity corresponds to the intersection of the full heliconical pitch of the TB phase with the fracture surface.

In quantitatively studying the TB phase with FFTEM, we must consider that certain fracture geometries will yield a measurement of the TB periodicity which may differ from the true helix pitch $p_H$, for instance when the helix axis makes an angle with the fracture plane or when the fracture plane is not normal to the incident electron beam (Fig. S17). Because the clean glass surfaces typically induce random planar anchoring in the TB phase and the FFTEM cells are thin (~3 – 10 μm) compared to the cell area (2 mm × 3 mm), the viewing direction **v** is almost always nearly perpendicular to the fracture plane **F**, meaning that the observed periodicity should correspond closely to the TB helix pitch. However, these geometrical effects contribute some anomalously small or large periodicities which have also been observed in other FFTEM studies of the TB phase [25,26]. We therefore observe a distribution of periodicities throughout the FFTEM replica, so that obtaining a representative set of images of the TB modulations is vitally important for a reliable sample set. With such a set of data, we can analyze the TB modulations statistically, considering all twist-bend domains that we observe and giving more weight to the modulations which are present over the larger areas of the fracture surface. Hence, we measure each observed periodicity by taking a spatial Fourier transform of the region and weight it by the domain area to create a weighted histogram distribution of the sample. A peak in this distribution represents a periodicity which is more likely to be observed during the FFTEM process, and because of the sample preparation, the pitch of the TB helix. Using this procedure, we can compare the FFTEM measurements in the TB phase in neat CB7CB and the various mixtures with RSoXS measurements.

Fig. S18 contains FFTEM histograms of mixtures of 5CB with CB7CB at various concentrations quenched ~10°C below the N–TB transition temperature, while the RSoXS data taken at the same temperature is laid on the x-axis and represented with a pink bar, its width indicating the range of pitches observed. These plots exhibit several blue bars grouped together into a peak which are centered at a periodicity which is slightly shorter than the pitch range measured by RSoXS. The RSoXS method is a more accurate and precise way to measure the heliconical pitch, although the statistical FFTEM method gives an excellent indication of the pitch length and its trend with increasing concentration of 5CB. Interestingly, the observed pitch distribution broadens dramatically in $x = 50.0$, with several prominent peaks at different values. At $x = 62.5$, we again obtain a grouped distribution of bars which is broad and centered on ~18 nm. The non-linear increase in the measured periodicity with 5CB concentration may signify some fundamental change in the twist-bend system or perhaps even a transformation to a unique liquid crystal structure with an ~18 nm modulation. Further investigation with RSoXS will be necessary to verify and determine the nature of this phenomenon.

We subsequently investigated the temperature dependence of a $x = 25.0$ mixture to discern any influence that the 5CB may have on the behavior of the TB pitch as a function of temperature (Fig. 18b). On quenching, just below the N–TB phase transition, we measure a very broad distribution of periodicities, with values ranging from ~7–14 nm. This behavior is similar to that observed for $x = 0$ (Fig. S18c and [1]). On quenching from further below the N–TB phase transition, we find that the periodicity distribution centers around ~8.4 nm consistently for several intermediate temperatures (5–25°C below $T_{NTB}$). At even lower quenching temperatures, we find the distribution of periodicities broadens again. This behavior may be due to phase separation of some of the 5CB from the mixture on quenching and disruption of the TB structure, though the optical texture appears homogenous before quenching.

We investigated the quenching temperature dependence of neat CB7CB by the statistical FFTEM method when quenched from five different temperatures (Fig. S18c) and compared this to the RSoXS method [1], which is an in-situ method and is the most reliable for temperature studies of the TB pitch. The FFTEM method contains a fast quenching step with immersion into liquid propane meant to freeze the sample from a given state into a glassy state which accurately represents the state of the sample before quenching. The FFTEM method does this well in less fluid liquid crystal phases, like the smectic phases of banana-shaped liquid crystals, for example [27,28], but gives less reliable results for the TB phase, which is a nematic-like fluid. In Fig. S18b, quenching from 1°C below the N–TB transition, we observe a broad distribution of periodicities from about 8 to 13 nm, demonstrating that the FFTEM measurement does not appear to capture the N–TB melting behavior which is represented by the small pink bar. On quenching from temperatures further below the N–TB phase, we see a persistent peak at 8 nm with a tail of histogram bars which becomes smaller on reducing the pre-quench temperature. We believe this tail encodes the differential response to dilation and compression observed in RSoXS, in which a variety of pitches are observed at elevated temperatures in the TB phase. When quenching from room temperature, we observe a single prominent peak at 8 nm in CB7CB, with the RSoXS data sitting almost precisely at the peak. This behavior seems to imply that the FFTEM method contains some of the temperature-dependent information which is gleaned from RSoXS, but that there is also some degree of annealing occurring, so that some TB domains relax and anneal to their low temperature pitch during the quenching process, which for CB7CB is 8 nm. The FFTEM statistical method is most sensitive to the pitch which is approached 'asymptotically' on cooling a TB-forming material. Temperature-dependent FFTEM studies can represent some temperature-dependent information which is convolved with several experimental effects which must be adequately considered and addressed to do quantitative studies, including geometrical effects, quenching effects, etc. The RSoXS experiments are preferable to temperature-dependent FFTEM experiments in general, but FFTEM can yield a surprisingly good measurement of the low temperature pitch to expected for a given TB-forming material.

Surprisingly, on quenching neat CB7CB from temperatures above the TB phase, we found we could still obtain large, uniformly aligned TB domains. This is a rather unexpected result, as our quench speed is very fast: ~100°C/sec. For reference, typical liquid crystal mesogens such as 5CB or MBBA supercool into glasses at much lower quench speeds of ~0.01–1°C/sec [29,30]. However, we even conducted the FFTEM experiment using copper planchettes, which are much more thermally conductive than glass to obtain a quench rate of ~10,000°C/sec [31], and still found large TB domains. This indicates that the formation of twist-bend domains in CB7CB is too quick to fully quench the isotropic phase. Based on our highest achievable quench speed and modelling the heat diffusion in our cell during the quench, we estimate that CB7CB forms micron-scale oriented twist-bend domains on a timescale of ~10 ms. This must be partially because the formation of TB domains only requires a collective reorientation of the molecules in the volume with minimal or no positional diffusion required, as opposed to the formation of a smectic phase from the isotropic melt, for instance, which requires collective reorientation and positional diffusion on the order of a molecular length in a brief time span.

The RSoXS pitch range as measured from the same temperature in FFTEM before quenching demonstrates the disagreement between RSoXS and FFTEM in the $x = 25.0$ mixture. It appears that the TB pitch anneals to a distribution roughly centered on 8.4 nm on quenching in this mixture at intermediate temperatures, indicating the 'asymptotically approached' pitch on cooling. The FFTEM study of neat CB7CB agrees better with the RSoXS data at the corresponding quench temperatures, although there is still evidence of the annealing behavior, represented as a peak centered at around 8 nm. The decrease in viscosity in the mixture may permit it to anneal more quickly into its most energetically favored heliconical pitch length, yielding the 'asymptotic' pitch at relatively higher temperatures than for CB7CB.

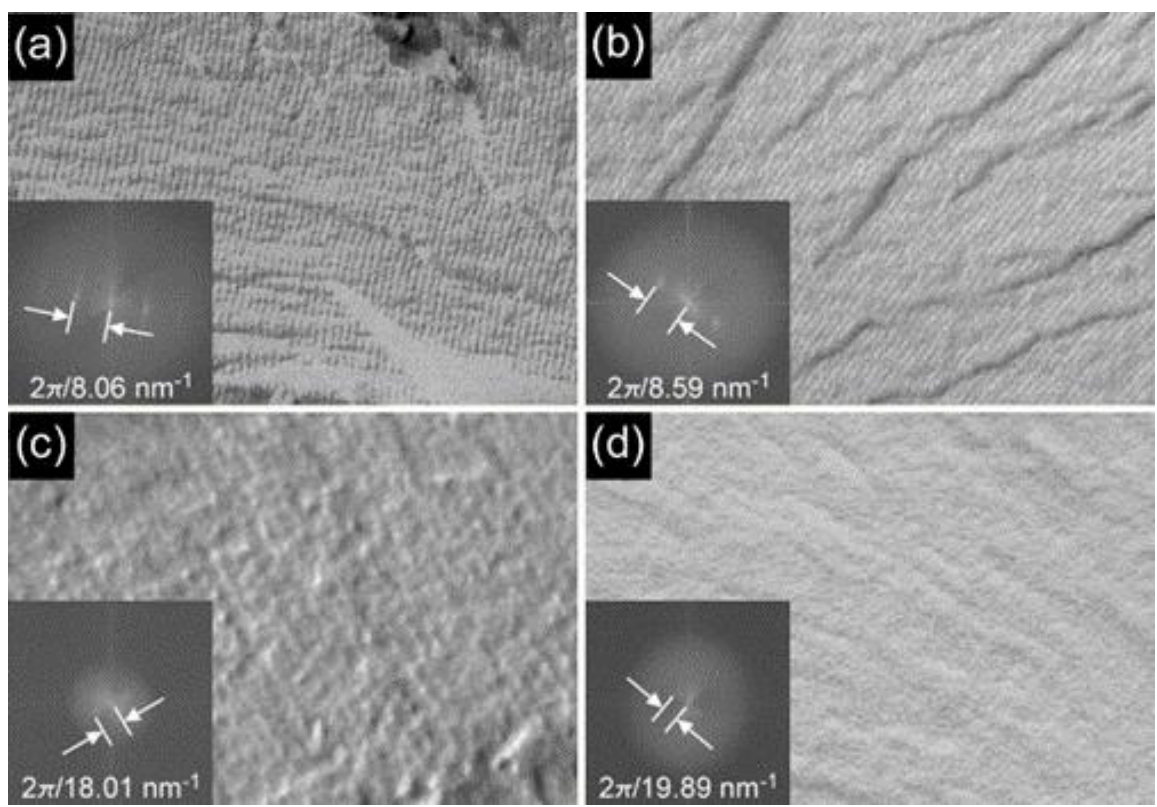

FIG. S17. FFTEM images showing twist-bend modulations in *a)* $x = 0$, *b)* $x = 25.0$, *c)* $x = 50.0$, and *d)* $x = 62.5$. The pitch is measured using spatial Fourier transforms, which are shown as insets. As the 5CB concentration increases in the mixtures, the twist-bend modulations are generally of larger spacing and smaller amplitude. All the images have the same scale.

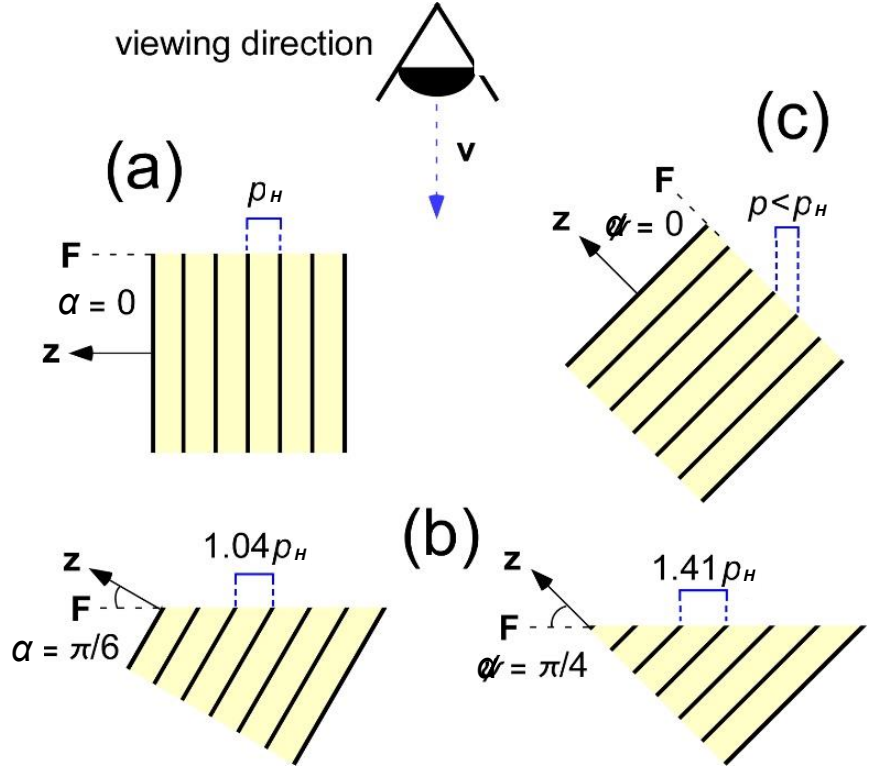

FIG. S18. Schematics of the possible configurations of twist-bend 'pseudo-layered' domains in FFTEM with helix axis **z** as viewed from viewing direction **v** onto fracture plane **F**, and considering the pseudo-layer spacing to be fixed at the value $p_H$. *a)* When **z** in the plane **F** and **v** $\perp$ **F**, we observe a periodicity corresponding to the ground state pitch $p_H$ of the TB phase. *b)* When **z** makes an arbitrary angle $\alpha$ with respect to the plane of **F** and **v** $\perp$ **F**, we observe periodicities $p > p_H$. *c)* When **z** // **F** and **v** makes an angle with respect to **F**, we observe a periodicity $p < p_H$.

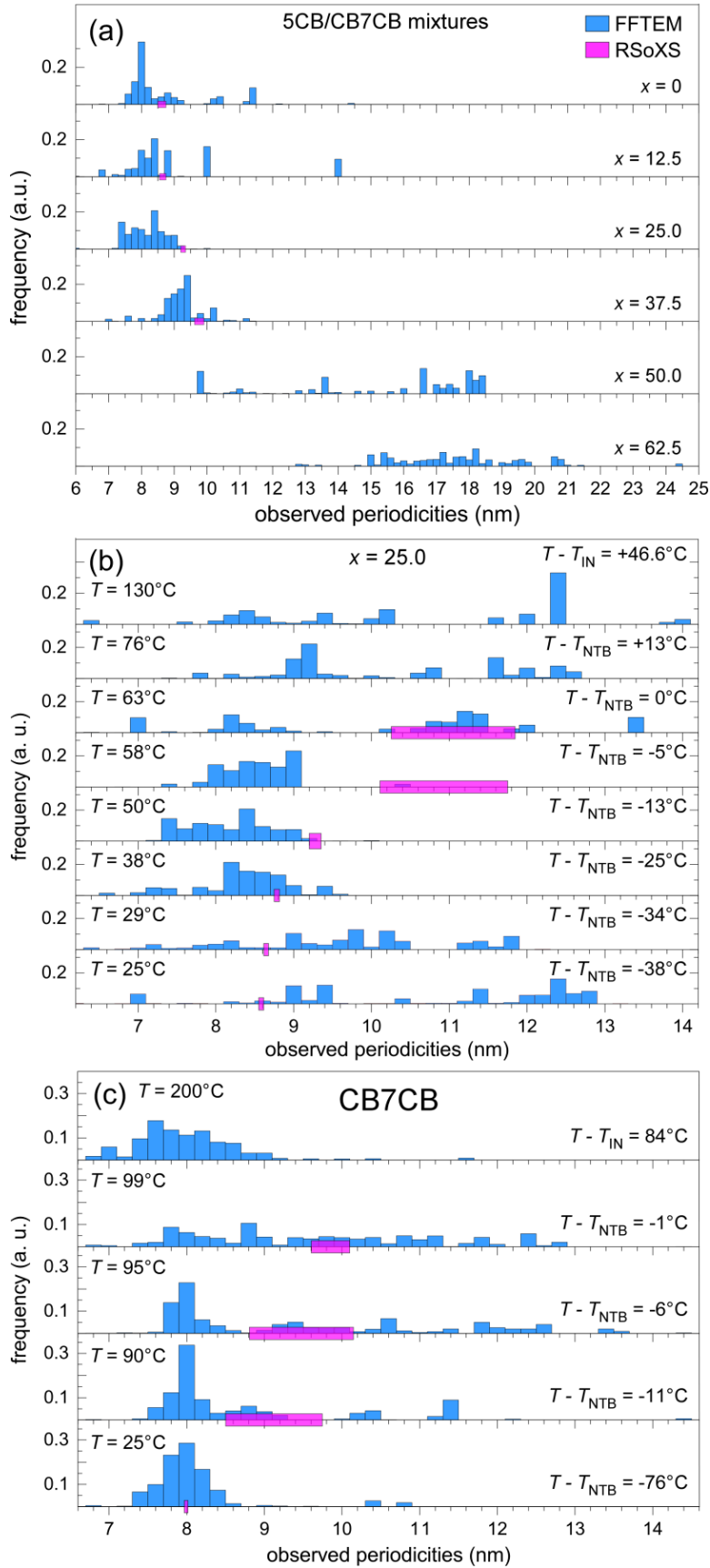

FIG. S19. Weighted measured periodicities of the 5CB/CB7CB mixtures and neat CB7CB. *a)* Distributions of the weighted periodicities of several 5CB/CB7CB mixtures indicate that the heliconical pitch increases steadily up to $x = 37.5$. RSoXS data overlaid onto the FFTEM distributions show roughly the same conclusion but are shifted slightly to longer pitch lengths. FFTEM measurements indicate some unusual and unexpected behavior for $x = 50.0$ and 62.5, in which the measured periodicity of these samples becomes very broad and the mean periodicity increases dramatically beyond those mixtures with lesser 5CB concentration. *b)* Distribution of periodicities in the $x = 25.0$ mixture as a function of quenching temperature, with the measured pitch range obtained by RSoXS at the given temperature overlaid as pink bars. *c)* Distribution of periodicities in neat CB7CB as a function of quenching temperature, with the measured pitch range obtained by RSoXS at the given temperature overlaid as a pink bar. Large, well-oriented TB domains were still observed at temperatures above the N–TB transition due to the quick formation of the TB compared to our minimum quenching time.